\documentclass[aps,twocolumn,superscript]{revtex4}

\usepackage{graphicx}

\usepackage[dvipsnames]{xcolor}
\usepackage{amsmath}
\usepackage{hyperref}
\usepackage{gensymb}
\hypersetup{colorlinks=true,
            linkcolor=blue,
            citecolor=blue,
            urlcolor=orange}

\begin{document}

\title{Exchange
interactions in itinerant magnets: the effects of local particle-hole irreducible vertex corrections and SU(2) symmetry of Hund interaction}
\author{A. A. Katanin}
\affiliation{Center for Photonics and 2D Materials, Moscow Institute of Physics and Technology, Institutsky lane 9, Dolgoprudny, 141700, Moscow region, Russia}
\affiliation{M. N. Mikheev Institute of Metal Physics of the Ural Branch of the Russian Academy of Sciences, S. Kovalevskaya Street 18, 620990 Yekaterinburg, Russia}

\begin{abstract}
We study exchange interactions in iron and nickel within the DFT+DMFT approach with density-density and SU(2) symmetric Coulomb interaction.  In particular, we analyze the representation of exchange interactions through the electron non-local Green's functions, connecting the local particle-hole (ph) irreducible interaction vertices. While the neglect of the local ph-irreducible vertex corrections, suggested previously within the dual fermion approach [E. A. Stepanov, et.al., Phys. Rev. Lett. {\bf 121}, 037204 (2018)], yields the result corresponding to the generalization of the magnetic force theorem approach, we argue that these vertex corrections are in fact less important for strong localized magnets, such as iron, but become more essential for weak itinerant magnets, such as nickel. At the same time, the overall account of the local vertex and self-energy corrections in the so-called renormalized magnetic force theorem  approach is essential for iron, and less important for nickel. 
The difference of the results of density-density and SU(2) symmetric Coulomb interaction is found to be relatively small, in contrast to the Curie temperature and the value of the local magnetic moment.
\end{abstract}

\maketitle

\section{Introduction}

The concept of exchange interactions, which originates from the consideration of interaction energy of many-electron atoms, was proposed by by P. Dirac, W. Heisenberg, and others in
1920's. Its extension to inter-atom interaction in transition metal compounds allowed W.~Heisenberg to explain their ferromagnetism, see Refs. \cite{goodenough,vonsovsky,mattis,yosida,white} for a review. This concept is natural for systems with constant on-site magnetic moments (i.e. insulators), but despite the number of  mechanisms of magnetic exchange proposed later beyond direct overlap of atomic orbitals, for itinerant systems, which have partly formed local magnetic moments, the exchange interactions are not uniquely defined. 

Lichtenstein et al. \cite{LKG1984,LKAG1987} proposed the formula for
magnetic exchange, based on infinitesimal rotations of magnetic moments (and magnetic
force theorem), which was intensively used for calculations of exchange
interactions \cite{KL2000,Kats2004,Solovyev2021,NNNFe1,NNNFe2,gFe}. However, the obtained exchange interactions characterize only low-energy excitations of the considered magnetic ground state. At the same time, at finite temperatures higher-energy states can be excited. Knowledge of the whole magnetic excitation spectrum can also be important for consideration of competition of various magnetic states, which can be close in energy but related by {\it finite} non-uniform rotation of spin degrees of freedom.

More comprehensive information on the magnetic excitation spectrum and, therefore, exchange interactions, can be obtained from the non-uniform magnetic susceptibility. The respective approach to the calculation of exchange interactions was proposed in Refs. \cite{chi,Antropov1} and justified within the `renormalized' magnetic force theorem \cite{Bruno}, see also Refs. \cite{Solovyev2021,Antropov,Kats2004}. These approaches use an RPA-type representation of susceptibility and express the exchange interactions through the inverse susceptibility, which (up to a constant) is equal to the inverse bare susceptibility.

However, in the presence of electronic correlations, which are accounted in the DFT approach only in average via the exchange energy correction, the susceptibility acquires self-energy and vertex corrections. While the self-energy corrections in realistic materials were actively studied  within the DFT+DMFT approach \cite{anisimov1997,LK1998,KL1999,DMFT_rev,DFTplusDMFT}, the non-local vertex corrections in multi-orbital systems have only started to be considered relatively recently, see, e.g., Refs. \cite{AbinitioDGA,AbinitioDGA1,Loon,OurJq,MyCo,MyCrO2}. Both, self-energy and vertex corrections have their impact on exchange interactions, reflecting the effect of correlations. Their effects are expected to be important in Hund metals \cite{Hund1,Hund2,Hund3,Hund4}, which have {\it long-lived} magnetic moments, and for which one can hope to obtain a physically meaningful definition of exchange interactions, which is similar to the one for insulators. Notably, such ``classical" magnets, as iron, nickel, and cobalt, belong to the class of Hund metals \cite{OurFe,OurGamma,Sangiovanni,AnisimovCo,MyCo}.

To account for the effect of correlations, the application of many-body techniques is required. The application of these techniques also gives the possibility of naturally describing the temperature dependence of exchange interactions. Recently, the dual fermion approach was used \cite%
{Stepanov1,Stepanov2,StepanovRev} to account for the effect of electronic correlations on exchange interactions. The resulting expression for the exchange interaction, similar to that, obtained from the magnetic force theorem, but containing local self-energy and vertex corrections, was derived. Obtaining this result amounts, however, to the neglect of some type of vertex corrections. 
These corrections, denoted in this paper as local particle-hole (ph) irreducible vertex corrections, correspond to the on-site interaction of particle-hole pairs, which propagate from one site to another, reflecting therefore their multiple scattering effect. 

In the present paper
we analyze the effect of the above mentioned local ph-irreducible vertex corrections, using the recently proposed formulation \cite{OurJq} based on consideration of inverse susceptibilities.
We show that while the vertex corrections are crucially important for local moment magnets, such as iron, the effect of the ph-irreducible local vertex corrections to the result of the leading order dual fermion approach is relatively small in this case. At the same time, the latter corrections provide a somewhat larger effect in systems with increased degree of itinerancy, for example in nickel, although the overall effect of vertex corrections for this material is weaker. 

We also consider the effect of SU(2) symmetric Coulomb interaction, since previous calculations of exchange interactions in realistic materials were performed for the density-density interaction (i.e. assuming Ising symmetry of Hund exchange). This allowed one for using segment impurity solvers, having smaller stochastic noise, but introduced additional approximation in the calculation of exchange interactions. In the present paper we show that the difference of the results for exchange interactions, assuming SU(2) symmetric and density-density Coulomb interaction,remains sufficiently small, at least for the considered systems with cubic lattice symmetry.

The plan of the paper is the following. In Sect. II we discuss the general formalism of calculation of exchange interactions in paramagnetic phase in DFT+DMFT approach and discuss the role of particle-hole irreducible local vertex corrections. In Sect. III we present numerical results for the density-density (Sect. IIIA) and SU(2) symmetric (Sect. IIIB) Coulomb interaction.  In Sect. IV we present Conclusions.

\vspace{-0.25cm}
\section{General formalism}

We consider the DFT+DMFT approach with the tight-binding multiorbital Hubbard model 
\begin{equation}
{H} = \sum_{{\bf k},\lambda\lambda',\sigma}H_{\bf k}^{\lambda\lambda'} c^+_{{\bf k}\lambda\sigma} c_{{\bf k}\lambda'\sigma}^{} +H_{\rm int}  - {H}_{\rm DC},
\end{equation}
describing electrons which are moved on the sites of the lattice ($\lambda,\lambda'$ are the orbital indices), subject to the on-site Coulomb repulsion $H_{\rm int}$, and $H_{\rm DC}$ is the double counting contribution, which is needed to keep the ab initio quasiparticle energies, obtained by diagonalizing $H_{\bf k}^{\lambda\lambda'}$ unchanged by the interaction. For the Coulomb repulsion we consider a general Hamiltonian
\begin{equation}
H_{\rm int}=\sum_{i,m_1 m_2 m_3 m_4,\sigma\sigma'} U_{m_1 m_2 m_3 m_4} c^{+}_{i m_1 \sigma} c^+_{i m_2 \sigma'} c_{i m_4 \sigma'} c_{i m_3 \sigma},
\end{equation} where $m_j$ are the orbital indices of the $d$ shell. The four-index matrix $U_{m_1,m_2,m_3,m_4}=\sum_{k=0}^4 F_k \alpha(l,k,m_1,m_2,m_3,m_4)$ is obtained from Slater representation, where $F_k$ are Slater's integrals, and $\alpha(...)$ are the angular Racah-Wigner numbers. In the case of Ising symmetry of Hund exchange, this interaction is reduced to the density-density form
\begin{equation}
H_{\rm int}=\frac{1}{2}\sum\limits_{i,mm',\sigma\sigma'} U^{mm^\prime}_{\sigma\sigma^\prime}
{n}_{im\sigma} {n}_{im^\prime\sigma^\prime},
\end{equation}  
where $n_{im\sigma}=c^+_{im\sigma} c_{im\sigma}$, and the two-index matrix elements $U^{mm'}_{\sigma\sigma'}$ are related to the four-index ones by $U^{m m'}_{\sigma,-\sigma} = U_{m m' m m'}$ and $U^{mm'}_{\sigma,\sigma}  = U^{mm'}_{\sigma,-\sigma} - U_{m m'm'm}$ .

We define exchange interactions in strongly correlated systems as described by the effective Heisenberg Hamiltonian for spin degrees of freedom%
\begin{equation}
H^{\mathrm{eff}}=-\frac{1}{2}\sum\limits_{\mathbf{q},rr^{\prime }}J_{%
\mathbf{q}}^{rr^{\prime }}\mathbf{S}_{\mathbf{q}}^r\mathbf{S}_{-\mathbf{q}%
}^{r^{\prime }},
\end{equation}%
where $r,r^{\prime }$ are the site indices in the unit cell; $\mathbf{q}$ is
the wave vector, $\mathbf{S}_{\mathbf{q}}^{r}$ are the Fourier transforms of
the static (classical) spins at the given sites of the lattice. To determine exchange interactions, we relate them to the non-uniform susceptibility using the RPA-type approach for spin degrees of freedom \cite%
{Izyumov} (cf. Refs.~%
\onlinecite{Antropov,chi,Antropov1,Otsuki,Bruno,IgoshevKatanin,BelozerovKat,Solovyev2021,OurJq,Izyumov})%
\begin{equation}
J_{\mathbf{q}}^{rr^{\prime }}=2\left[ (\chi ^{r}_{\mathrm{loc}})^{-1}\delta
_{rr^{\prime }}-(\chi _{\mathbf{q}})_{rr^{\prime }}^{-1}\right], 
\label{Jqdef}
\end{equation}%
where $\chi _{\mathbf{q}}^{rr^{\prime }}=2\langle\langle S^{z,r}_{\mathbf{q}} | S^{z,r}_{-\mathbf{q}}\rangle\rangle_{\omega=0}$ and $\chi ^{r}_{\mathrm{loc}}=2\langle\langle S^{z,r}_{i} | S^{z,r}_{i}\rangle\rangle_{\omega=0}$ are
the orbital-summed non-local and local static susceptibilities ($S^{z,r}_{i}$ being the Fourier transform of $S^{z,r}_{\mathbf{q}}$) \cite{RefNote}, and in
the second term in the right hand side of Eq. (\ref{Jqdef}) the matrix inversion with respect to
the site indices in the unit cell is assumed. We emphasize that Eq. (\ref{Jqdef}) treats only the bilinear component of magnetic exchange; to obtain
higher-order exchange interactions (biquadratic, ring exchange, etc.) one would need to consider correlation functions of four and more spin operators. 

\begin{figure*}[t]
\centering
\includegraphics[width=0.7\linewidth]{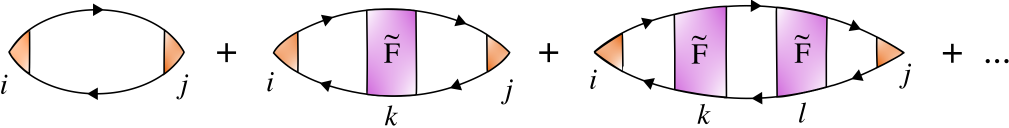}
\caption{(Color online) Diagrammatic representation of Eq. (\ref{Jqmain}) in real space ($i,j,k,l$ are spatial indices which include the cell index and the site index $r$ within the unit cell); orbital indices are omitted for brevity. Solid lines correspond to the non-local Green's functions $\widetilde G_{ij,\nu}$ (which are the Fourier transforms of $\widetilde G_{{\mathbf k},\nu}^{rr'}$), left (right) orange triangles represent the local vertices $\widetilde{\Lambda}_\nu$ ($\widetilde{\Lambda}^T_\nu$) , violet parts correspond to the local ph irreducible local interaction $\widetilde F^r_{\nu\nu'}$}  
\label{Fig:chiF}
\end{figure*}

In dynamical mean field theory the non-local susceptibility $\chi_{\mathbf q}$ is represented as a sum of the ladder diagrams, constructed from the electron Green's functions $G_{\mathbf{k}%
,\nu }^{rm,r'm^{\prime }}$, with the particle-hole irreducible local vertices connected by the bubbles (describing propagation of particle-hole pairs) $\chi _{\mathbf{q},\nu }^{0,rm,r^{\prime }m^{\prime }}=-T\sum_{\mathbf{%
k}}G_{\mathbf{k},\nu }^{rm,r'm^{\prime }}G_{\mathbf{k+q},\nu
}^{r'm^{\prime },mr}$ \cite{DMFT_rev,OurJq}, $\nu$ denote fermionic Matsubara frequencies. To derive the decomposition of exchange interaction into the non-local bubbles
\begin{equation}
\widetilde{\chi }_{\mathbf{q,}\nu}^
{rm,r'm^{\prime }}=-T\sum_{\mathbf{k}}\widetilde{G}_{\mathbf{k},\nu }^{rm,r'm^{\prime
}}\widetilde{G}_{\mathbf{k+q},\nu }^{r'm^{\prime },rm},
\label{chi0Nonloc}
\end{equation}
constructed from the non-local Green's functions $\widetilde{G}_{\mathbf{k},\nu }^{rm,r'm^{\prime }}=G_{\mathbf{k}%
,\nu }^{rm,r'm^{\prime }}-G_{\mathrm{loc},\nu }^{rm}\delta_{mm'}\delta_{rr'}$ (for simplicity, we assume the local Green's function $G_{\mathrm{loc},\nu }^{rm}=\sum_{\mathbf k} G_{\mathbf{k}%
,\nu }^{rm,rm}$ diagonal in the orbital indices)
and the local particle-hole irreducible vertex corrections, we use the ``dual'' representation of the non-uniform susceptibility \cite{chiDMFT,Loon}. Using matrix notations with respect to site-, orbital- and frequency indices the orbital-summed non-uniform static susceptibility in the ladder approximation reads \cite{Loon}
\begin{align}
\chi _{\mathbf{q}}^{r r^{\prime }}
=\chi ^{r}_{\mathrm{loc}}\delta_{rr'}+\left\{
\Lambda \widetilde{\chi }_{\mathbf{q}}\left[ I-F_{\mathrm{loc}}\widetilde{%
\chi }_{\mathbf{q}}\right] ^{-1}\Lambda ^{T}\right\} _{rr^{\prime }},
\label{chiq}
\end{align}%
where $F^{r,m m^{\prime }}_{\mathrm{loc},\nu\nu'}$ are the full local interaction vertices, obtained from the two-particle correlation functions, 
$\Lambda ^{rm}_\nu =(G_{\mathrm{loc},\nu}^{rm})^{-2}\langle c_{rm\sigma,\nu}^{+}c_{rm\sigma,\nu
}S_{\omega =0}^{z,r}\rangle$ is the local triangular vertex, $\Lambda^{T,mr}_\nu=\Lambda^{rm}_\nu$ and matrix multiplications with respect to the site, orbital, and frequency indices are assumed (the quantities depending on a single site-, orbital-, or frequency index are considered as diagonal in that index). As it is discussed in Ref. \cite{Loon}, using Eq. (\ref{chiq}) allows reducing the number of the considered fermionic Matsubara frequencies in the vertex calculation (we typically use $N_\nu=60$ positive and negative Matsubara frequencies). Rearranging Eq. (\ref{chiq}) and combining it with Eq. (\ref{Jqdef}), we obtain an identical expression for the exchange interactions (\ref{Jqdef})
in the form (see Appendix A)
\begin{align}
J_{\mathbf{q}}^{rr^{\prime }} &=2\sum_{\nu \nu ^{\prime },r^{\prime \prime
},mm^{\prime }m^{\prime \prime }}\widetilde{\Lambda }^{rm}_\nu \widetilde{%
\chi }_{\mathbf{q,}\nu }^{rm,r^{\prime \prime }m^{\prime \prime }}  \left[\vphantom{\sum_{m'''}}
\delta _{r^{\prime \prime }m''\nu,r^{\prime }m'\nu'}
\right.   \label{Jqmain} \\
& -\left.\sum_{m'''}\widetilde{F}^{r^{\prime \prime },m^{\prime \prime }m'''}_{\mathrm{loc},\nu\nu'}\widetilde{\chi }_{\mathbf{q},\nu ^{\prime
}}^{r^{\prime \prime }m^{\prime \prime \prime},r^{\prime }m^{\prime }}\right]
_{r^{\prime \prime }m^{\prime \prime }\nu ,r^{\prime }m^{\prime }\nu
^{\prime }}^{-1}\widetilde{\Lambda }_{\nu ^{\prime
}}^{T,m'r'}  \nonumber
\end{align}%
where $\widetilde{\Lambda }^{rm}_\nu =(\chi
^{r}_{\mathrm{loc}})^{-1}\Lambda ^{r m}_\nu  $ is the amputated triangular vertex, $%
\widetilde{F}^{r,mm^{\prime }}_{\mathrm{loc},\nu \nu ^{\prime }}=F^{r,mm^{\prime }}_{\mathrm{loc},\nu \nu ^{\prime }}-\widetilde{\Lambda }^{T,rm}_\nu\chi ^{r}_{\mathrm{loc}}\widetilde{\Lambda }^{rm^{\prime }}_{\nu ^{\prime
}}$ is the object that we refer to in the following as the local {ph-irreducible} vertex, and $\delta_{ab}$ assumes pairwise equality of constituents of $a$ and $b$. 
We note that the Eq. (\ref{Jqmain}) remains valid also for orbital-resolved exchange interactions if instead of the site indices $r$ we consider multi-indices ${\rm r}=(r,m_r)$, containing also the index of the orbital. In that case, the susceptibilities $\chi ^{\mathrm {rr'}}_{\mathrm{loc}}\equiv\chi ^{r,m_r m_r'}_{\mathrm{loc}}\delta_{rr'}$ and the triangular vertices $\Lambda ^{\mathrm {rr}'}_\nu\equiv\Lambda ^{r,m_r m_r'}_{\nu}\delta_{rr'}$ become matrices with respect to the site-orbital indices, since the operator $S_{\omega}^{z,\mathrm{r}}\equiv S_{\omega}^{z,r,m_r}$ is attributed to a certain orbital $m_r$, and multiplication by $\chi ^{\mathrm {rr'}}_{\mathrm{loc}}$ (as well as its inverse) and $\Lambda^{\mathrm {rr'}}_\nu$ (and its transpose) should be understood as matrix multiplication. We also note that the result for the inverse susceptibility, corresponding to the Eq. (\ref{Jqmain}) in the single-orbital case, was derived previously within the dual fermion approach in Ref. \cite{Stepanov2} (see Supplemental Material of that paper). Our derivation generalizes this result to the multi-orbital case, and, at the same time, it is more straightforward, since it is based on the ladder result for the non-local susceptibility, which has a clear diagrammatic representation.

Diagrammatic form of the Eq. (\ref{Jqmain}) in real space is shown in Fig. \ref{Fig:chiF}. In its full form, Eq. (\ref{Jqmain}) accounts for both local self-energy and
vertex corrections. This equation allows us to unify various approaches to the calculation of exchange interactions. In the leading order in $\widetilde{\chi }_{q}$, or
neglecting the local {ph-irreducible} vertex $\widetilde{F}^{r,mm^{\prime }}_{\mathrm{loc},\nu \nu ^{\prime }}$ (i.e approximating 
$F^{r,mm^{\prime }}_{\mathrm{loc},\nu \nu ^{\prime }}\simeq\widetilde{\Lambda }^{T,r,m}_\nu\chi ^{r}_{\mathrm{loc}}\widetilde{\Lambda }^{r,m^{\prime }}_{\nu ^{\prime
}}$, as in Refs \cite{Stepanov1,Stepanov2}, corresponding to the first diagram in Fig. \ref{Fig:chiF}), the Eq. (\ref{Jqmain}) reduces to the  generalization of the magnetic force theorem result (see also Refs. \cite{Stepanov1,Stepanov2,StepanovRev})%
\begin{equation}
J_{\mathbf{q}}^{rr'}=2\sum_{\nu ,mm^{\prime }}\widetilde{\Lambda }^{rm}_\nu \widetilde{%
\chi }_{\mathbf{q,}\nu }^{rm,r^{\prime}m^{\prime }}
\widetilde{\Lambda }_{\nu }^{T,r'm'},
\label{Jq0}
\end{equation}%
containing dynamic vertices $\widetilde{\Lambda }^{rm}_{\nu }$, as well as the self-energy effects. Eq. (\ref{Jq0}) is {analogous} to that derived in Refs. \cite{Stepanov1,Stepanov2,StepanovRev,OurJq}, but uses triangular vertices for orbital-summed magnetic moment (yielding also exchange interaction for orbital-summed magnetic moments), instead of the orbital-resolved ones. 
The magnetic force theorem result corresponds to choosing frequency-independent $\widetilde{\Lambda }^{rm}_{\nu }$, determined by the spin splitting in the magnetically ordered phase, see also Refs. \cite{Stepanov1,Stepanov2,StepanovRev}.

Let us also discuss the relation of the present approach (Eq. (\ref{Jqmain})) to the `renormalized' magnetic force theorem approach \cite{Antropov1,Bruno, Antropov}. In our notations the latter corresponds to the 
full neglect of the local vertex corrections, $F_{\mathrm{loc}%
}=0,$ which implies the summation of the whole series in Fig.~\ref{Fig:chiF} with $\widetilde{\Lambda }^{rm}_{\nu }=(\chi ^{r}_{\mathrm{loc,0}})^{-1}$, $\widetilde F^{r,mm'}_{\nu\nu'}=-\widetilde{\Lambda }^{T,r,m}_\nu\chi ^{r}_{\mathrm{loc}}\widetilde{\Lambda }^{r,m^{\prime }}_{\nu ^{\prime}}=- \widetilde{\Lambda }^{rm}_{\nu }$ and $\chi ^{r}_{\mathrm{loc}}=\chi^{r}_{0,%
\mathrm{loc}}\equiv -T\sum_{{\nu },mm'}G_{\mathrm{loc},\nu }^{r,mm^{\prime
}} G_{\mathrm{loc,}\nu }^{r,m^{\prime }m}$. Substituting this into
Eq. (\ref{Jqmain}), we obtain%
\begin{align}
J_{\mathbf{q}}^{rr^{\prime }}&=2\sum_{r''}(\chi ^{r}_{\mathrm{loc,0}})^{-1}\widetilde{\chi }_{\mathbf{q} }^{rr^{\prime \prime}}\left( \chi^{r''}_{\mathrm{loc,0}}\delta_{r''r'}+
\widetilde{\chi }_{\mathbf{q} }^{r''r^{\prime }}\right) ^{-1}\notag\\&=2\left[ (\chi^{r}_{\mathrm{loc,0}%
})^{-1}\delta_{rr'}-\left( \chi^{r}_{\mathrm{loc,0}}\delta_{rr'}+
\widetilde{\chi }_{\mathbf{q} }^{rr^{\prime }}\right) ^{-1}%
\right]   \label{BA}
\end{align}%
where $\widetilde{\chi }_{\mathbf{q}}^{rr^{\prime }}=\sum_{\mathbf{\nu ,}mm^{\prime }}%
\widetilde{\chi }_{\mathbf{q,}\nu }^{rm,r^{\prime }m'}$ for the orbital-summed exchange interactions and the replacements $r\rightarrow \mathrm{r}$, $r'\rightarrow \mathrm{r}'$ with $\widetilde{\chi }_{\mathbf{q}}^{\mathrm{rr}^{\prime }}=
\sum_{\mathbf{\nu}}\widetilde{\chi }_{\mathbf{q,}\nu }^{\mathrm{rr}^{\prime }}$ is used for orbital-resolved interactions.
When neglecting also the frequency-dependent self-energy corrections (keeping only Hartree contribution) and considering orbital-resolved susceptibilities, Eq. (\ref{BA}) coincides with that proposed originally in Refs. \cite{Antropov1,Bruno,Antropov}. At the same time, as it follows from the Eq. (\ref{Jqmain}), the contribution $\widetilde{\Lambda }^{T,r,m}_\nu\chi ^{r}_{\mathrm{loc}}\widetilde{\Lambda }^{r,m^{\prime }}_{\nu ^{\prime
}}$ is aimed at the subtraction of the local ph-reducible part from the full local vertex $F^{r,mm'}_{{\rm loc},\nu\nu'}$. Therefore, account of the vertex corrections is essential for the 'renormalized' approach, as we also demonstrate explicitly in the numerical analysis below.


We note that the approaches (\ref{Jqmain}), (\ref{Jq0}), and (\ref{BA}), 
being all derived from Eq. (\ref{Jqdef}), correspond to the \textit{bilinear }%
exchange interactions. This is in contrast to the consideration of Ref. \cite%
{StepanovRev}, which related the difference between Eqs. (\ref{Jqdef}) and (%
\ref{Jq0}) to including the effect of multi-spin exchange in the former
equation. The above presented derivation implies that Eq. (\ref{Jq0}) is
just an approximation to the Eqs. (\ref{Jqdef}) and (\ref{Jqmain}). Its advantage is that it requires only the knowledge of the triangular local vertices, which calculation in the impurity solver requires less computational cost. At the same time,
as one can see from Eq. (\ref{Jqmain}), the corrections to the equation (\ref%
{Jq0}) occur as a result of multiple local interactions $%
\widetilde{F}_{\mathrm{loc}}$ of the particle-hole pairs, which propagate
with the interactions at the other sites of the lattice. Because of the
approximation $F_{\mathrm{loc}}\simeq \widetilde{\Lambda }^{T}\chi _{\mathrm{%
loc}}\widetilde{\Lambda }$, these contributions were neglected in the dual fermion approach of Refs. 
\cite{Stepanov1,Stepanov2,StepanovRev}. At the same time, the `renormalized magnetic force theorem' approach of Refs. \cite{Antropov1,Bruno,Antropov}  corresponds to the neglect of full interaction vertices, and requires only knowledge of the electronic self-energy. As we will see in the following Section, the irreducible interaction $\widetilde{F}_{\mathrm{loc}}$ is in general smaller than the full local interaction ${F}_{\mathrm{loc}}$ due to subtraction of the reducible part, and at least for strong magnets, it produces the corrections, which are not too large. At the same time, for weak magnets neglecting full vertex $F$ 
provides reasonable results, although with an overestimate of exchange interactions (see next Section).

For the single-orbital model in the localized limit $t\ll U$ of the
single-band Hubbard model Eq. (\ref{Jq0}) reproduces the standard $4t^{2}/U$ nearest-neighbor antiferromagnetic exchange (see Appendix B). In this respect, the corrections, represented by an expansion of the denominator in Eq. (\ref{Jqmain}), contribute to
higher orders in $t^{2}/U$, which also produce longer-range exchange interactions. These corrections become essential in itinerant
systems with $U\sim t$. The role of these corrections in the multi-orbital case is analyzed in the next Section.

\section{Numerical results}

For obtaining numerical results we consider $\alpha$-iron and nickel, as strongly correlated materials. The band structures are obtained using the Quantum Espresso package \cite{QE} with ultrasoft pseudopotentials from the SSSP PBEsol Precision library \cite{PPP} and $16\times 16\times16$ momentum grid, with subsequent Wannier projection of 3d, 4s, and 4p states with maximally localized Wannier functions, obtained within Wannier90 package \cite{Wannier90}. As in the previous study \cite{OurJq}, we consider the paramagnetic phase. We use the Slater parametrization of Coulomb interaction with $U=F^0=4$~eV and Hund exchange $J=0.9$~eV and around-mean-field form of the double counting correction \cite{AMF}. We also use the corrections for finite frequency box in Bethe-Salpeter equations \cite{My_BS}; further details can be found in Ref. \cite{OurJq}. Similarly to Ref. \cite{OurJq}, we subtract the local part (which appears because of the approximations used, but remains typically small) from the obtained exchange interactions.


\subsection{Dynamical mean field theory for the density-density interaction}


For DMFT calculations in the case of density-density interaction (i.e. Ising symmetry of Hund exchange) we use segment continuous time Monte-Carlo (CT-QMC) \cite{CT-QMC} iQIST solver \cite{iQIST}, which allows us to considerably reduce the stochastic noise and obtain the details of the frequency dependence of the vertex functions over a sufficiently broad frequency range. 

\begin{figure}[b]
\centering
\includegraphics[width=0.85\linewidth]{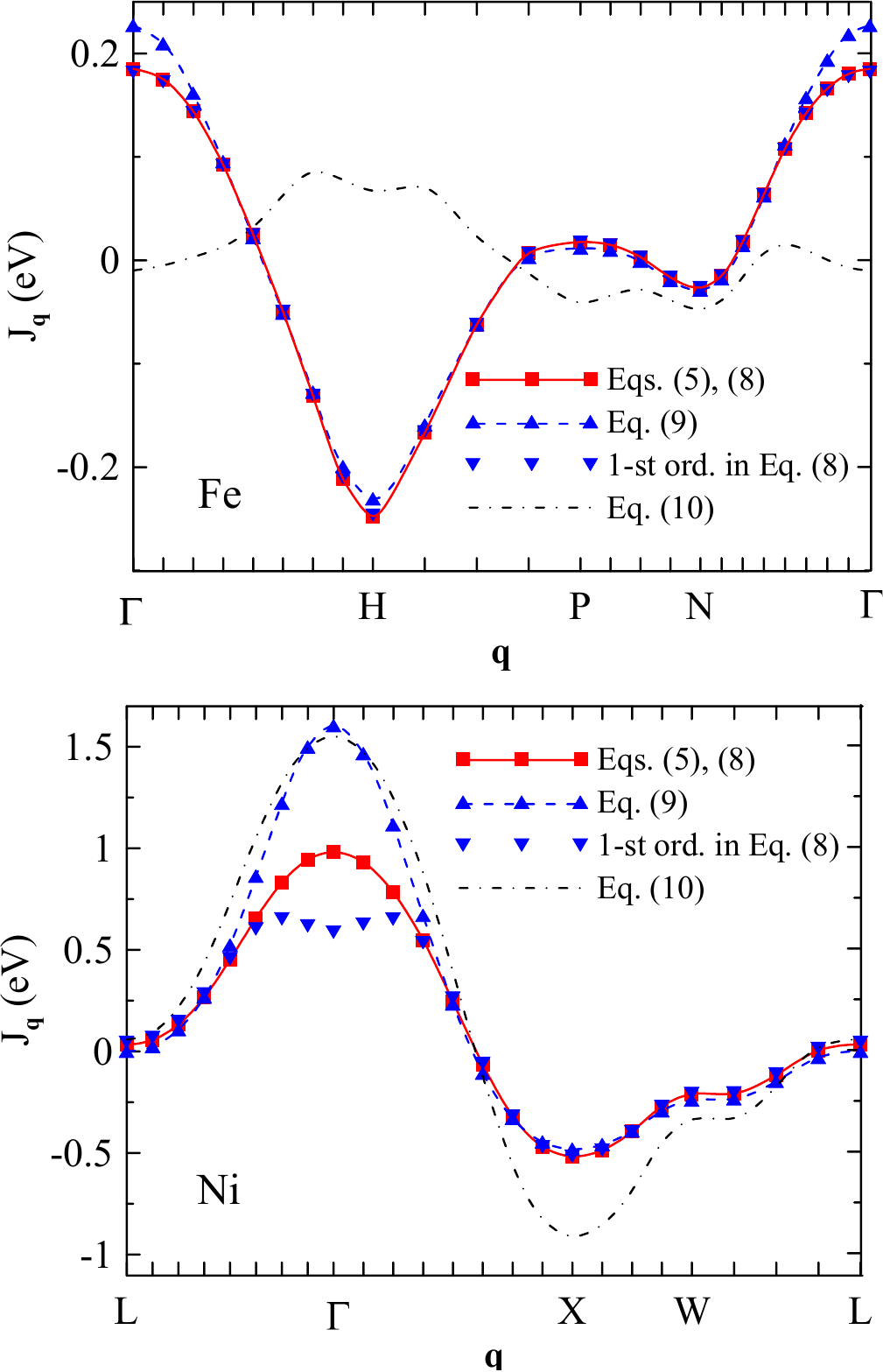}
\caption{(Color online) The momentum dependencies of exchange interactions in density-density approximation at $\beta=10$~eV$^{-1}$. Solid line (squares) corresponds to the ladder approximation, Eq. (\ref{Jqdef}) (or, equivalently, Eq. (\ref{Jqmain})), dashed line (upward triangles) corresponds to the neglect of the local ph irreducible vertex, Eq. (\ref{Jq0}), downward triangles show the first order correction to the Eq. (\ref{Jq0}) (i.e. expanding Eq. (\ref{Jqmain}) to linear order in $\widetilde{F}$), dash-dotted lines correspond to the Eq. (\ref{BA}). The inset corresponds to the exchange interaction $J_{\mathbf q}$ in nickel for $\beta=5$~eV$^{-1}$.}  
\label{Fig:Jq}
\end{figure}

\begin{figure*}[t]
\centering
\includegraphics[width=0.9\linewidth]{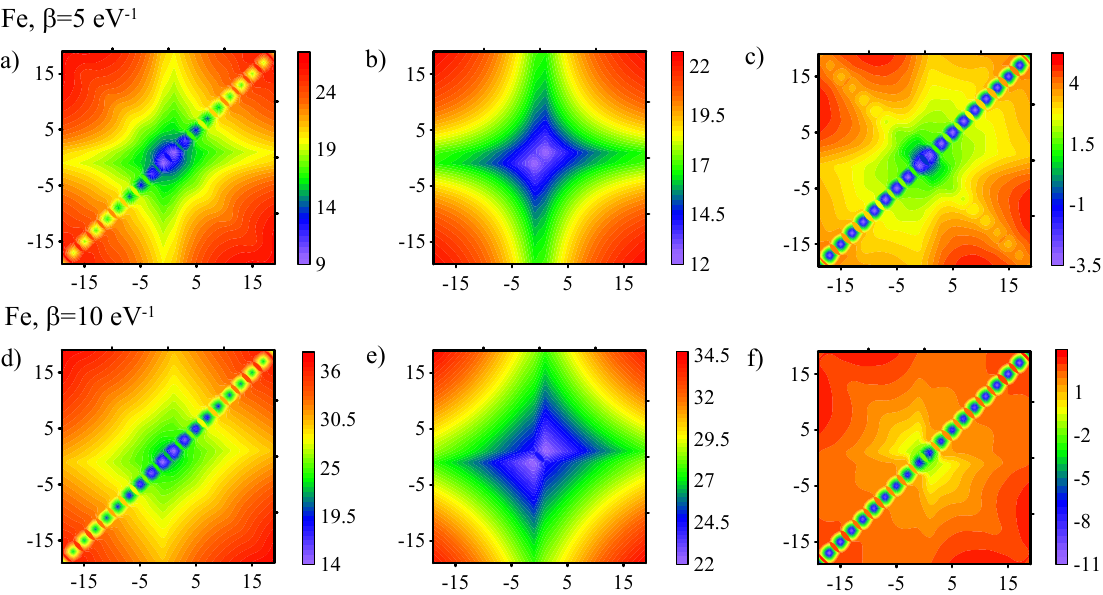}
\includegraphics[width=0.9\linewidth]{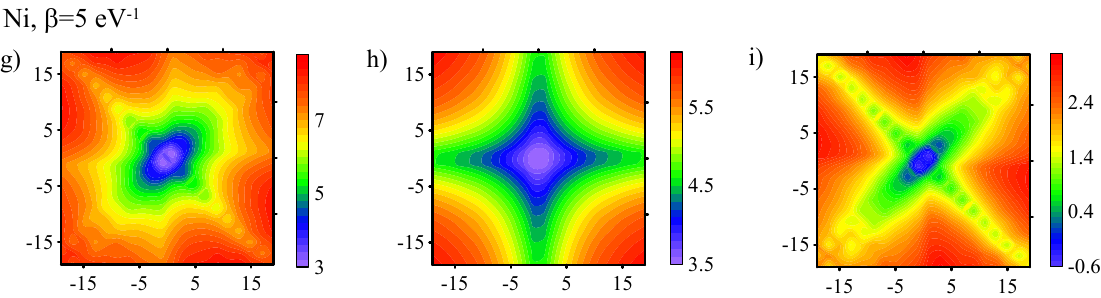}
\caption{(Color online) Contour plots of the frequency $\nu_n$ and $\nu_n'$ dependence of real part of local interaction vertices in iron (a-f) and nickel (g-i) at $\beta=5$~eV$^{-1}$ (a-c,g-i) and $\beta=10$~eV$^{-1}$ (d-f): (a,d,g) Full vertex $F^{mm}_{\nu\nu'}$, (b,e,h) Local ph reducible part $\widetilde \Lambda^m_{\nu}\chi_{\rm loc}\widetilde\Lambda^m_{\nu'}$, and (c,f,i) local ph irreducible part $\widetilde{F}^{mm}_{\nu\nu'}$ for $m=d_{xz}$ (the interactions of the other orbitals look similarly). The numbers at the axes correspond to $\nu/(\pi T)$ and $\nu'/(\pi T)$. Scale bars are in units of eV. }  
\label{Fig:F_Fe}
\end{figure*}

In Fig. \ref{Fig:Jq} we plot the momentum dependencies of exchange interactions. The obtained exchange interactions are close to those previously obtained in Ref. \cite{OurJq}, small differences are attributed to different downfolding procedure (maximally localized Wannier orbitals in the present study vs. localized Wannier states of the previous study), and using the orbital-summed triangular local vertices in the present study, which better reflect the formation of local magnetic moments among different orbitals. 

In iron, we find only a small difference between the exchange interaction neglecting the ph-irreducible local interaction, Eq. (\ref{Jq0}), and the corresponding result within the ladder approximation, Eq. (\ref{Jqmain}). Therefore, in this case the local ph-irreducible vertex corrections, related to the vertex $\widetilde{F}^{\rm loc}$, are small. At the same time, as it is discussed in Sect. II, the vertex corrections themselves (related to the full local vertex $F$) are extremely important. Neglecting these corrections yields even the absence of ferromagnetism at not too low temperatures, since the resulting magnetic susceptibility $\widetilde{\chi}_{\mathbf q}$, as well as the respective magnetic exchange, is peaked near the wave vector ${\mathbf q}_H=(0,0,2\pi)/a$ and symmetry related points. Expanding the matrix inverse in Eq. (\ref{Jqmain}) to the first order in the ph-irreducible local vertex $\widetilde{F}$ (denoted as the first-order result in Fig. \ref{Fig:Jq}) leads to the result close to zeroth-order approximation, showing therefore that multiple particle-hole scattering effects are responsible for the difference of the results (\ref{Jqmain}) and (\ref{Jq0}). 

In nickel, we find a larger difference between the results of Eqs. (\ref{Jq0}) and (\ref{Jqmain}), at the point $\Gamma$, than in iron. The difference constitutes almost $\sim 50\%$ of exchange interaction (\ref{Jqmain}) at the considered temperature, and further increases with a decrease of temperature (see next Subsection). In this case, however, full neglect of vertex corrections (i.e. using the Eq. (\ref{BA})) produces a more reasonable result, than in iron. Also, in the case of nickel, the first-order approximation somewhat overestimates the correction to the exchange interactions near the point $q=0$, which shows also that multiple particle-hole scattering effects are important near this wave vector. 

\begin{figure}[b]
\centering
\includegraphics[width=0.75\linewidth]{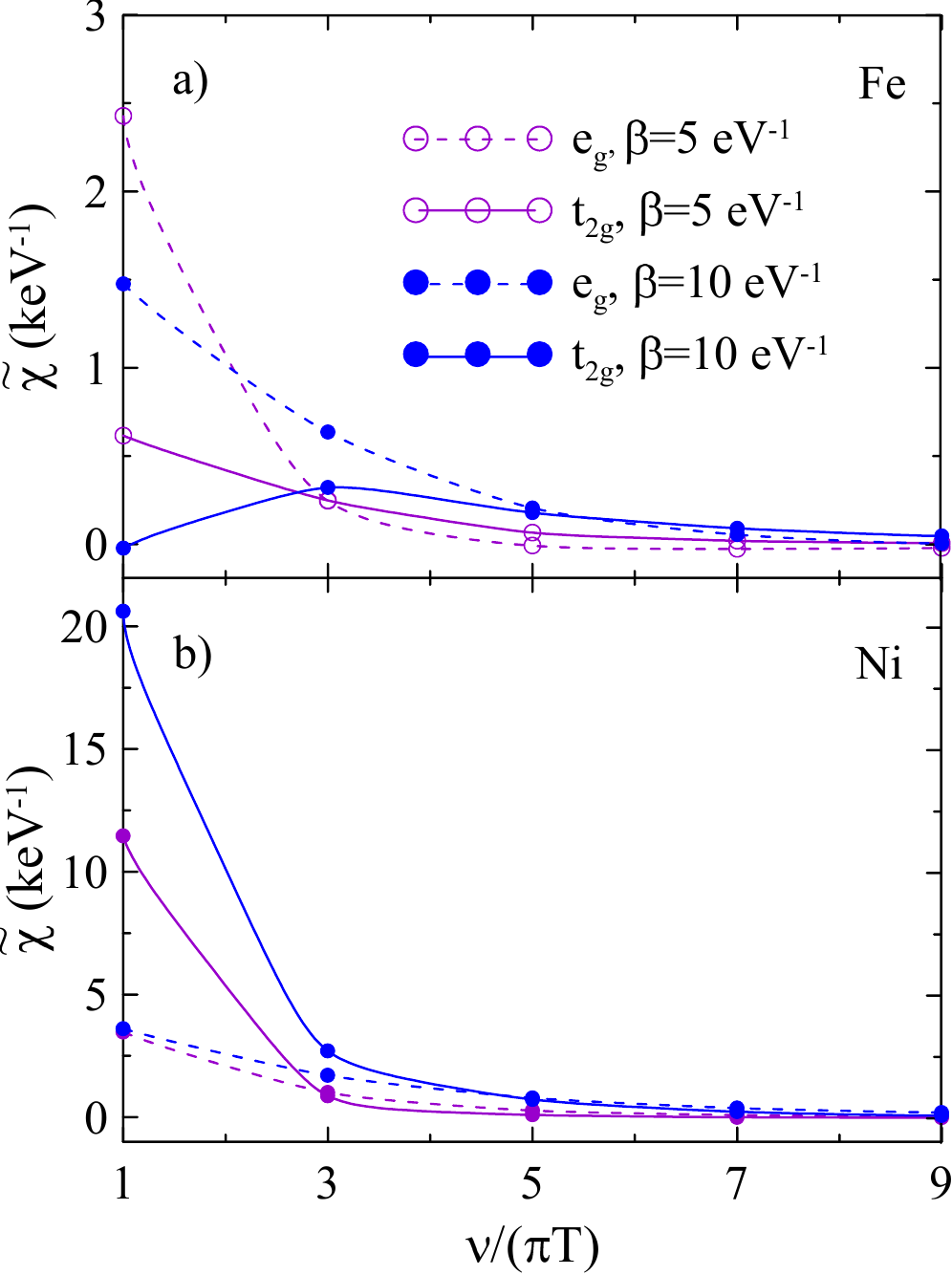}
\caption{The frequency dependence of the bubble $\widetilde{\chi }_{\mathbf{q=0,}\nu }^{mm}$ for $m=t_{2g}$ (solid lines) and $m=e_g$ (dashed lines) orbitals in iron (a) and nickel (b) at $\beta=5$~eV$^{-1}$ (open circles) and $\beta=10$~eV$^{-1}$ (filled circles).}  
\label{Fig:chi1}
\end{figure}

To understand the origin of relative smallness of local ph-irreducible vertex corrections, in Fig. \ref{Fig:F_Fe} we show the frequency dependence of interaction vertices in iron and nickel for one of the $t_{2g}$ orbitals (the interaction vertices for the other combinations of orbitals look similarly). One can see that the full local vertex $F^{mm}_{\nu\nu'}$ is rather large in iron ($\sim 20$~eV at $\beta=5$~eV$^{-1}$ and $\sim 30$~eV at $\beta=10$~eV$^{-1}$), and somewhat smaller in nickel ($\sim 10$~eV), where it is only weakly temperature dependent. However, to a large extent these contributions originate from the local ph reducible part. The absolute value of the local ph irreducible part is substantially smaller ($\lesssim 5$~eV) for both materials, except $\nu=\nu'$ in iron, where the vertex reaches somewhat larger absolute values with decreasing temperature. While the full vertex has both, vertical, horizontal ($\nu=0$ or $\nu'=0$), and diagonal ($\nu=\pm\nu')$  peculiarities (see Fig. \ref{Fig:F_Fe}a,d,g), they originate from different contributions. The vertical and horizontal peculiarities originate from the local ph reducible part (Fig. \ref{Fig:F_Fe}b,e,h), and correspond to peculiarities of triangular vertices. At the same time, diagonal $\nu=\pm \nu'$ contributions correspond to the contributions of the crossed particle-hole channel ($\nu=\nu'$) and particle-particle channel ($\nu=-\nu'$) to the particle-hole irreducible part (Fig. \ref{Fig:F_Fe}c,f,i). The particle-particle contribution is less pronounced in iron, and more pronounced in nickel.

In spite of rather localized electron motion in iron, the effect of the local ph irreducible interaction is further suppressed by the smallness of the non-local bubbles (\ref{chi0Nonloc}). 
The frequency dependence of the non-local part of the polarization bubble $\widetilde{\chi }_{\mathbf{q=0,}\nu }^{mm}$ is shown in Fig. \ref{Fig:chi1} (we omit the site indices $r$ for 1 site in the unit cell). One can see that this bubble is rather small in iron ($\sim {\mathrm{keV}}^{-1}$), with the largest contribution from $e_g$ orbitals, related to the body-centered cubic crystal structure (see, e.g., Ref. \cite{OurFe}). The bubble decays fast with frequency, such that at the considered not too low temperatures only the contribution of the lowest Matsubara frequencies dominates. Because of the smallness of the non-local bubble, the contribution of the denominator in Eq. (\ref{Jqmain}) is also relatively small. In nickel, we find a somewhat larger non-local part of the bubble $\widetilde{\chi }_{\mathbf{q=0,}\nu }^{mm}$ due to less localized electron behavior, mainly from t$_{2g}$ orbitals, related to face-centered cubic crystal structure. Correspondingly, the contribution of the ph-irreducible local interaction is larger. Moreover, the temperature dependence of the bubble $\widetilde{\chi }_{\mathbf{q=0,}\nu }^{mm}$ is qualitatively different in iron and nickel.  In iron the non-local bubble is suppressed with decrease of temperature due to the increase of quasiparticle damping \cite{OurFe,Sangiovanni,BelozerovKat}. This suppression of the bubble partly compensates the increase of the absolute value of ph-irreducible interaction vertices, and the contribution of these vertices to exchange interactions does not grow fast with decrease of temperature (see next subsection). On the other hand, in nickel the scattering rate decreases with decrease of temperature, which yields {\it increase} of the non-local bubble, which corresponds to increasing itinerancy. Respectively, as we show in the next subsection, the effect of the ph irreducible local interaction increases fast with decrease of temperature.  





Therefore, we find that 
the importance of the ph-irreducible
local interaction $\widetilde{F}$ increases with increasing the degree of itinerancy of the system.
Since we expect that the effects of interaction will be weakened in the symmetry-broken phase, the validity of the result of the magnetic force theorem therefore relies on (i) smallness and fast decay in the frequency axis of the non-local vs. full polarization bubble (ii) relative smallness of local ph irreducible vertex vs. full interaction vertex. At the same time, the renormalized magnetic force theorem approach requires, in general, consideration of vertex corrections.


\subsection{SU(2) invariant Coulomb interaction}

\begin{figure}[t]
\centering
\includegraphics[width=0.85\linewidth]{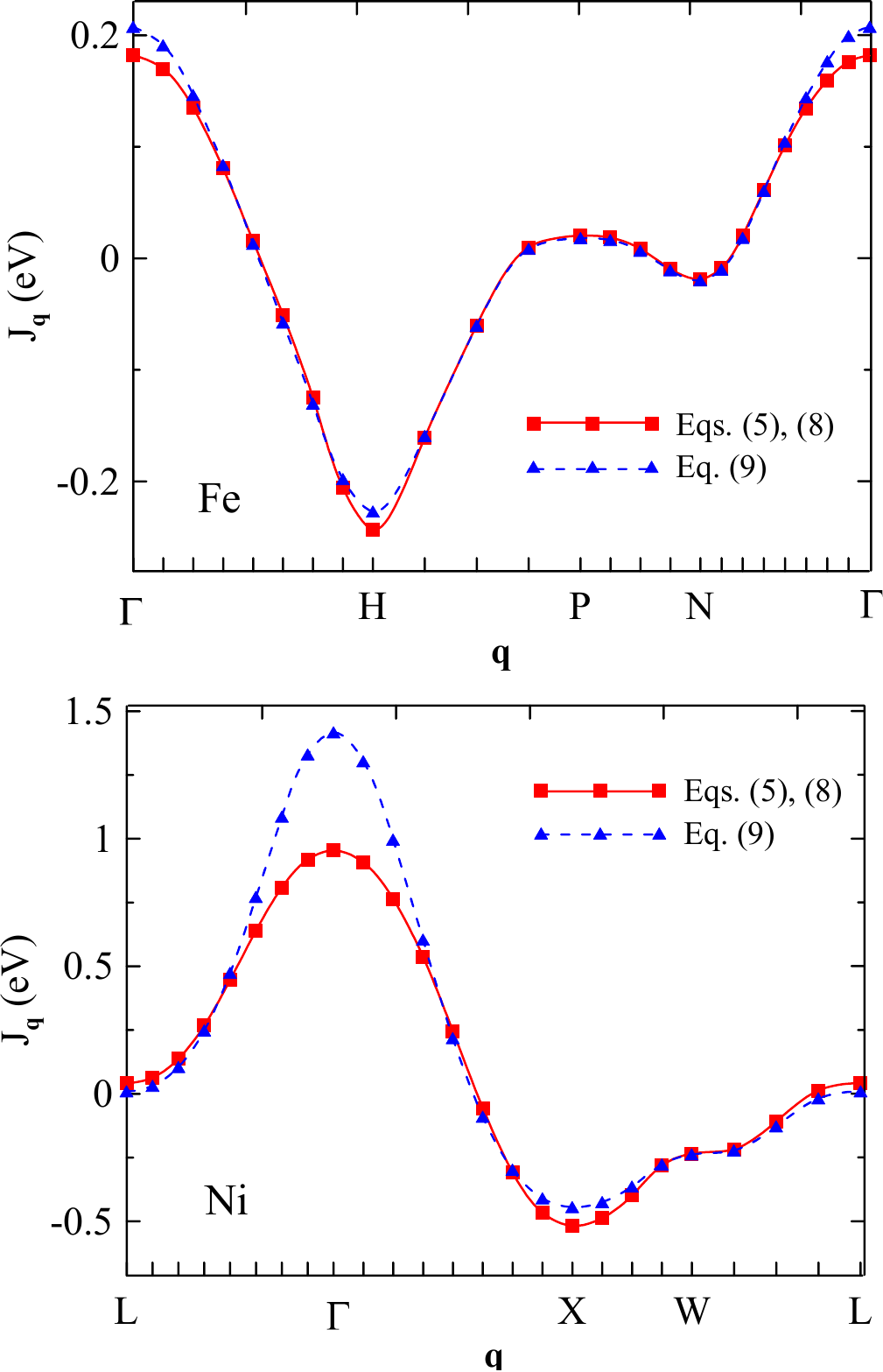}
\caption{(Color online) The momentum dependencies of exchange interactions in SU(2) symmetric approach at $\beta=10$~eV$^{-1}$. Solid line (squares) corresponds to the Eq. (\ref{Jqdef}), dashed line (upward triangles) corresponds to the lowest order approximation with respect to the local ph irreducible vertex, Eq. (\ref{Jq0}).}  
\label{Fig:JqSU2}
\end{figure}

In this subsection we show that the obtained results remain qualitatively unchanged with account of full SU(2) symmetry of Coulomb interaction. To treat this interaction, we use the TRIQS \cite{triqs} CT-HYB \cite{ct-hyb} impurity solver, combined with TRIQS DFTTools package \cite{DFTTools}. The non-local susceptibilities and exchange interactions are calculated using the Eqs. (\ref{Jqdef})-(\ref{Jqmain}) (the results are also compared to those from Eq.  (\ref{Jq0})), where the interaction vertices and non-local bubbles are considered depending on two orbital indices, in accordance with the assumption of diagonal self-energy and local Green's functions with respect to the orbital indices. The resulting momentum dependence of exchange interactions is presented in Fig. \ref{Fig:JqSU2}. One can see that for iron the deviation of the result (\ref{Jqmain}) from the results of the leading-order approximation (\ref{Jq0}) is even smaller than for the density-density interaction. For nickel, similarly to the density-density interaction, we find a stronger difference between the leading order and ladder approximation. 

\begin{figure}[t]
\centering
\includegraphics[width=0.85\linewidth]{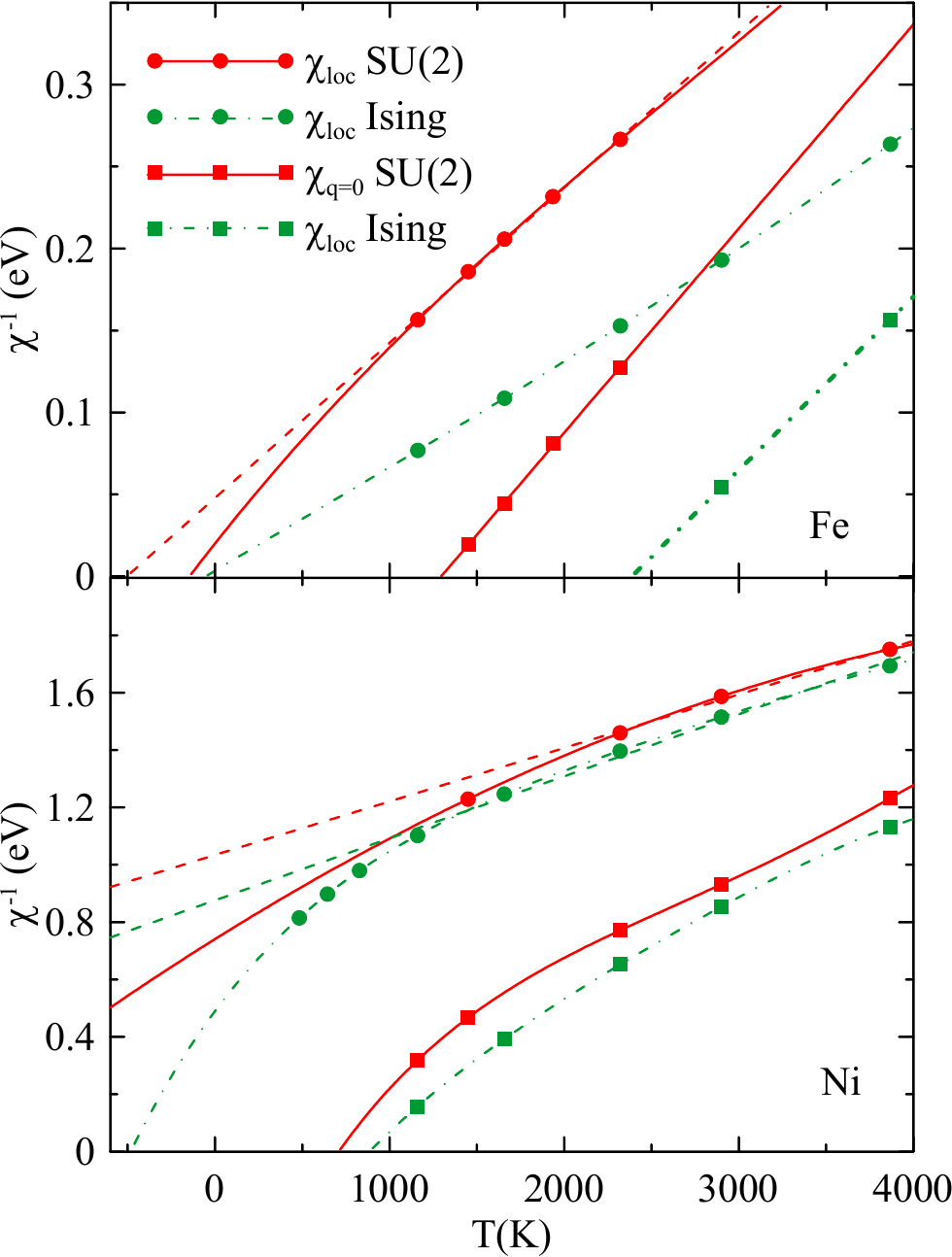}
\caption{(Color online) The temperature dependence of inverse uniform (squares) and local (circles) susceptibilities in iron and nickel. Solid red lines correspond to SU(2) symmetric Coulomb interaction, dash-dotted green lines corresponds to the density-density interaction (Ising symmetry of Hund exchange). The extrapolations to the low-temperature regime are obtained by polynomial fits; dashed lines show the result of the linear extrapolation for comparison.}  
\label{Fig:chi}
\end{figure}

\begin{figure}[t]
\centering
\includegraphics[width=0.85\linewidth]{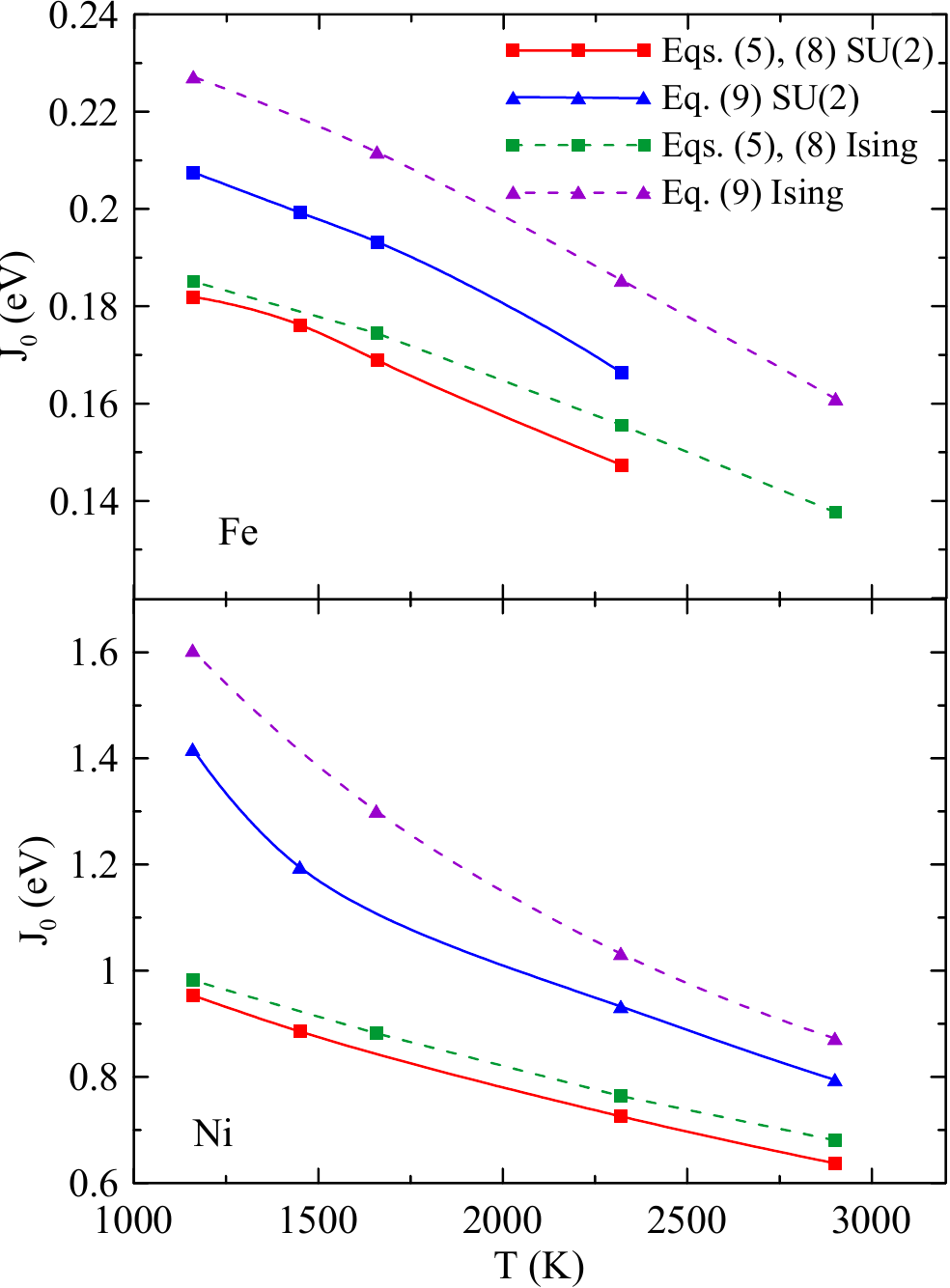}
\caption{(Color online) The temperature dependence of exchange interactions in iron and nickel. Solid (dashed) lines correspond to SU(2) symmetric (density-density with the Ising symmetry of Hund exchange) interaction. Squares corresponds to the Eq. (\ref{Jqdef}), dashed line (upward triangles) corresponds to the lowest order approximation with respect to the local ph irreducible vertex, Eq. (\ref{Jq0}). }  
\label{Fig:JT}
\end{figure}

In Fig. \ref{Fig:chi} we present the obtained temperature dependencies of inverse susceptibilities. In agreement with the previous studies \cite{Sangiovanni,AnisimovSU2} the Curie temperature and magnetic moments are suppressed for SU(2) symmetry of Hund exchange. For iron, we obtain $T_C=1280$~K, which is much smaller than the Curie temperature in the density-density approximation $T_C=2380$~K, and much closer to the experimental value $T_C\simeq 1043$~K. The remaining difference to the experimental Curie temperature occurs as a result of spin fluctuations beyond the considered mean-field approach (see below). Also, the value of local magnetic moment, obtained from the slope of inverse local magnetic susceptibility, $\mu^2_{\rm loc}=10.9\mu_B^2$ in SU(2) symmetric approach is much closer to the experimental data ($\mu^2\simeq 10\mu_B^2$), than the corresponding value $\mu^2_{\rm loc}=16.1\mu_B^2$ for density-density interaction. Interestingly, the Kondo temperature of iron, obtained from the vanishing linear extrapolation from the considered temperature range of inverse local susceptibility, $\chi_{\rm loc}^{-1}(-\sqrt{2}T_K)=0$ (see, e.g., Refs. \cite{Kondo1,Kondo2,Kondo3,Kondo4,Kondo5}), is much larger than for the case of density-density interaction, see also Ref. \onlinecite{Sangiovanni}. At the same time, using polynomial fit for extrapolation, which accounts for the change of the slope of inverse susceptibility at low temperatures, yields a much smaller Kondo temperature $T_K\simeq 100$~K, comparable to that for the density-density interaction. This stresses a larger degree of itinerancy of iron in SU(2) symmetric approach, corresponding to a smaller local magnetic moment than in the density-density approximation. Using the obtained exchange interactions and the above-mentioned values of local magnetic moments, we also estimate corrections to the DFT+DMFT Curie temperatures within RPA approach (see, e.g., Ref. \cite{MyCo} for the details). This way, we obtain $T_C\simeq 1800$~K for the density-density Coulomb interaction and $T_C \simeq 950$~K for SU(2) Coulomb interaction, the latter result is in good agreement with the experimental data. 

In nickel, we find non-linear in temperature inverse uniform and local susceptibilities for both, SU(2) symmetric and density-density interactions. The downturn of the local and uniform susceptibilities is related to the peculiarity of the band structure: the edge of the $e_g$ band, located near the Fermi level (see Fig. 3d of Ref. \cite{Sangiovanni}). This suppresses magnetic moment in Ni at low temperatures, yielding larger slope of inverse susceptibility. In case of nickel, the difference between the results for density-density and rotation-invariant form of Coulomb interaction is not so pronounced. Yet, the DMFT Curie temperature for SU(2) case $T_C=710$~K, which is also closer to the experimental value ($T_C=615$~K), than $T_C=880$~K for the density-density interaction. The reliable determination of the Kondo temperature, low-temperature magnetic moments, and respective corrections to the DMFT magnetic transition temperatures of nickel requires consideration of local susceptibility at low temperatures \cite{Sangiovanni}, which we do not perform in the present paper.

The temperature dependence of exchange interactions in SU(2) symmetric approach is presented and compared to the results of the density-density approximation in Fig. \ref{Fig:JT}. As it was anticipated in Ref. \cite{OurJq}, because of the partial compensation of the decrease of Curie temperatures and local magnetic moments, the difference of exchange interactions for SU(2) symmetric and density-density interactions is not pronounced. We find $J_0\simeq 0.19$~eV for iron and $J_0\simeq 1$~eV in nickel at low temperatures, which is close to the values obtained within the density-density interaction.

\section{Conclusion}

In conclusion, we have analyzed the effect of the local particle-hole irreducible vertex corrections, which describe multiple scattering of particle-hole pairs at different lattice sites, on exchange interactions of strongly correlated magnets. In iron, we find a minor effect of this type of vertex corrections, which describe deviation from the result, generalizing the magnetic force theorem approach to the paramagnetic phase. Yet, we find the full vertex corrections themselves are quite important in the case of iron, changing strongly the momentum dependence of exchange interactions.

On the contrary, in nickel, despite the relatively smallness of vertex corrections, the particle-hole irreducible vertex corrections appear to be more important and constitute 25-50\% of the exchange interaction in the considered temperature range.

Considering SU(2) invariant Coulomb interaction, we find that in agreement with the previous studies \cite{Sangiovanni,AnisimovSU2} it strongly affects Curie temperatures and the size of magnetic moments, but has a much weaker effect on exchange interactions, which shows that the density-density approximation is a good starting point for studying these interactions.

For further studies, the extension of the consideration of the present paper to the symmetry broken (ferromagnetic) phase is of certain interest, since it may allow one to describe a crossover to the low-temperature regime, where correlation effects are weakened by magnetic splitting of the electronic spectrum, and also study the temperature evolution of the dispersion and damping of magnons. Another interesting aspect is the consideration of frustrated and/or low-dimensional magnetic systems, having stronger magnetic fluctuation effects and/or lower symmetry of the crystal field.

\section*{Acknowledgements} 
The author is deeply indebted to the late Prof. Yu. N. Gornostyrev for his  contribution to this field and grateful to E. Stepanov, I. Solovyev, and E. Agapov for discussions. Performing the DMFT calculations was supported by the Russian Science Foundation
(project 19-72-30043-P). DFT calculations were carried out within the framework of the state assignment of the Ministry of Science and Higher Education of the Russian Federation for the IMP UB RAS. The calculations were
performed on the cluster of the Laboratory of Material
Computer Design of MIPT.

\appendix

\section{Derivation of Eq. (\protect\ref{Jqmain})}

In this appendix we use Eqs. (\ref{Jqdef}) and (\ref{chiq}) 
considered as matrix relations. In particular, the triangular vertices $\Lambda^{rm}_{\nu }=(G_{\mathrm{loc},\nu}^ {rm})^{-2}\langle c_{r\nu m}^{+}c_{r\nu
m}S_{\omega =0}^{z}\rangle$ and $\Lambda^{{\mathrm r}m}_{\nu }=(G_{\mathrm{loc},\nu}^ {rm})^{-2}\langle c_{r\nu m}^{+}c_{r\nu
m}S_{\omega =0}^{z,n}\rangle$, determining the orbital-summed and orbital-resolved interactions (${\mathrm r}=(r,n)$, $n$ is the orbital index) are considered as 
(diagonal with respect to site indices) $N_{r}\times N_{s}N_{%
\mathrm{orb}}N_{\nu }$ matrices ($N_{s}$ is the number of sites in the unit
cell, $N_{\mathrm{orb}}$ is the number of correlated orbitals per site, $%
N_{\nu }$ is the number of fermionic Matsubara frequencies, considered in
the vertex calculation); $N_r=N_s$ for the orbital-summed interactions and $N_r=N_s N_{\rm orb}$ for the orbital-resolved interactions, respectively. We also consider $\chi ^{rr^{\prime }}_{%
\mathrm{loc}}=\chi ^{r}_{\mathrm{loc}}\delta _{rr^{\prime }}$ as the $N_r\times N_r$ matrix, and the interaction $F_{\mathrm{loc},\nu
\nu ^{\prime }}^{rm,r'm'}=F_{\mathrm{loc},\nu
\nu ^{\prime }}^{r,mm'}\delta _{rr^{\prime }}$ as $N_{s}N_{\mathrm{orb}}N_{\nu
}\times N_{s}N_{\mathrm{orb}}N_{\nu }$ matrix. 

Using
matrix notations, let us introduce right- and left-inverse matrices 
\begin{equation}
\Lambda \cdot \Lambda ^{-1}=(\Lambda ^{T})^{-1}\cdot \Lambda ^{T}=E,
\end{equation}
which are defined as%
\begin{align}
\Lambda ^{-1} &=\Lambda ^{T}(\Lambda \Lambda ^{T})^{-1}, \notag\\
(\Lambda ^{T})^{-1} &=(\Lambda \Lambda ^{T})^{-1}\Lambda, 
\end{align}%
and $E$ is the unity $N_{r}\times N_{r}$ matrix. Using these definitions, we
find%
\begin{widetext}
\vspace{-0.5cm}
\begin{align}
J_{\mathbf{q}}/2 &=\chi _{\mathrm{loc}}^{-1}-\left\{ \chi _{\mathrm{loc}%
}+\Lambda \widetilde{\chi }_{\mathbf{q}}\left[ I-F_{\mathrm{loc}}\widetilde{%
\chi }_{\mathbf{q}}\right] ^{-1}\Lambda ^{T}\right\} ^{-1}  \nonumber \\
&=\chi _{\mathrm{loc}}^{-1}-\left\{ E+\chi _{\mathrm{loc}}^{-1}\Lambda 
\widetilde{\chi }_{\mathbf{q}}\left[ I-F_{\mathrm{loc}}\widetilde{\chi }_{%
\mathbf{q}}\right] ^{-1}\Lambda ^{T}\right\} ^{-1}\chi _{\mathrm{loc}}^{-1}
\nonumber \\
&=\chi _{\mathrm{loc}}^{-1}-\left\{ (\Lambda ^{T})^{-1}\left[ I-F_{%
\mathrm{loc}}\widetilde{\chi }_{\mathbf{q}}+\Lambda ^{T}\chi _{\mathrm{loc}%
}^{-1}\Lambda \widetilde{\chi }_{\mathbf{q}}\right] \left[ I-F_{\mathrm{loc}%
}\widetilde{\chi }_{\mathbf{q}}\right] ^{-1}\Lambda ^{T}\right\} ^{-1}\chi_{\mathrm{loc}}^{-1}  \nonumber \\
&=\chi _{\mathrm{loc}}^{-1}-(\Lambda ^{T})^{-1}\left[ I-F_{\mathrm{loc}}%
\widetilde{\chi }_{\mathbf{q}}\right] \left[ I-F_{\mathrm{loc}}\widetilde{%
\chi }_{\mathbf{q}}+\Lambda ^{T}\chi _{\mathrm{loc}}^{-1}\Lambda 
\widetilde{\chi }_{\mathbf{q}}\right] ^{-1}\Lambda^{T}\chi_{\mathrm{loc}}^{-1}  \nonumber \\
&=\chi _{\mathrm{loc}}^{-1}\Lambda \widetilde{\chi }_{\mathbf{q}}%
\left[ I-\left( F_{\mathrm{loc}}-\Lambda ^{T}\chi _{\mathrm{loc}%
}^{-1}\Lambda \right) \widetilde{\chi }_{\mathbf{q}}\right] ^{-1}
\Lambda ^{T}\chi _{\mathrm{loc}}^{-1}  \nonumber \\
&=\widetilde{\Lambda } \widetilde{\chi }_{\mathbf{q}}\left[ I-\left(
F_{\mathrm{loc}}-\widetilde{\Lambda }^{T}\chi _{\mathrm{loc}}\widetilde{%
\Lambda }\right) \widetilde{\chi }_{\mathbf{q}}\right] ^{-1} \widetilde{%
\Lambda }^{T}  \label{Jqder}
\end{align}%
\end{widetext}
where $I$ is the unity $N_{s}N_{\mathrm{orb}}N_{\nu
}\times N_{s}N_{\mathrm{orb}}N_{\nu }$ matrix, and $\widetilde{\Lambda }=\chi _{\rm loc}^{-1}\Lambda $. Appearance of the
series in Eq. (\ref{Jqder}) is directly related to the series considered in
RPA formula of Eq. (\ref{Jqdef}). Written explicitly with site, orbital, and
frequency indices, this yields Eq. (\ref{Jqmain}) for the orbital-summed interaction. The same
result with the replacements $r\rightarrow{\mathrm{r}}$ and $r'\rightarrow{\mathrm{r}'}$ holds for the orbital-resolved exchange interactions.

\section{$t/U$ expansion of exchange interactions in the single-band Hubbard model}

In this Appendix we consider the expansion of exchange interaction in powers of $t/U$  in the half-filled single-band Hubbard model with nearest-neighbor hopping $t_{ij}=t$ and on-site Coulomb repulsion $U$. To this end, we consider the expansion of the corresponding electron Green's function in real space
\begin{align}
G_{ij,\nu}&=(i\nu+U/2-t_{ij}-\Sigma_\nu)^{-1}\\&=G_\nu\delta_{ij}+G_{\nu }^{2}t_{ij}+G_{\nu }^{3}t_{ik}t_{kj}+G_{\nu
}^{3}t_{ik}t_{kl}t_{lj}+...\notag
\end{align}%
where we use atomic limit of the self-energy $\Sigma_\nu=U/2+U^2/(4i\nu)$,  $G_{\nu }=({i\nu+U/2 -\Sigma_\nu})^{-1}$
is the respective local Green's function (the applicability of the atomic limit of the self-energy is discussed below), and summation over repeated indices is assumed. We further use the atomic limit of the interaction vertex at $T\ll U$ and zero bosonic frequency \cite{Hafermann,ToschiV}
\begin{align}
F_{{\rm loc},\nu \nu ^{\prime }}&=U-\frac{U^{3}}{4}\left( \frac{1}{\nu ^{2}}+\frac{1}{%
\text{$\nu ^{\prime }$}^{2}}\right) 
+\frac{U^{2}}{4T}\left( 1+\frac{U^{2}}{4\nu ^{2}}\right) ^{2}\delta _{\nu
\nu ^{\prime }}\notag \\
&-\frac{3U^{5}}{16\nu ^{2}\nu ^{\prime 2}}+\frac{U^{2}}{2T}\left( 1+\frac{U^{2}}{4\nu ^{2}}\right)
\left( 1+\frac{U^{2}}{4\nu ^{\prime 2}}\right). \label{Fat}
\end{align}%
Performing the respective summations over frequencies (considering $T\rightarrow 0$ limit and replacing frequency sums by integrations in the absence of discontinuities), we obtain 
\begin{equation}
{\Lambda }_{\nu }=1-T \sum_{i\nu'} F_{{\rm loc},\nu \nu'}G_{\nu'}^2=1-\frac{U^2}{4\nu^2}+ 
\frac{U}{2T}\left( 1+\frac{U^{2}}{4\nu ^{2}}\right) \label{Lat}
\end{equation}%
and
\begin{equation}
\chi _{\text{loc}}=-\frac{1}{2\pi}\int\limits_{-\infty}^{\infty}\Lambda_{\nu}G_{\nu}^2{d\nu}=\frac{1}{2T}.
\end{equation}%
In the following, we keep only the leading in the limit $T\rightarrow 0$ (i.e. last in the Eq. (\ref{Lat})) term in $\Lambda_\nu$. In this case the second term in $\widetilde{F}_{\mathrm{loc},\nu \nu ^{\prime }}=F_{\mathrm{loc},\nu \nu ^{\prime }}-{\Lambda }_\nu\chi^{-1} _{\mathrm{loc}}{\Lambda }_{\nu ^{\prime
}}$ exactly cancels the last term of the vertex $F_{{\rm loc},\nu \nu'}$ in Eq. (\ref{Fat}) and we obtain
\begin{align}
\widetilde{\Lambda}_\nu&=U\left( 1+\frac{U^{2}}{4\nu ^{2}}\right),\\
\widetilde{F}_{{\rm loc},\nu \nu ^{\prime }}&=U-\frac{U^{3}}{4}\left( \frac{1}{\nu ^{2}}%
+\frac{1}{\text{$\nu ^{\prime }$}^{2}}\right) +\frac{U^{2}}{4T}\left( 1+\frac{U^{2}}{4\nu ^{2}}\right)
^{2}\delta _{\nu \nu ^{\prime }}\notag\\&-\frac{3U^{5}}{16\nu ^{2}\nu
^{\prime 2}}.
\end{align}
Remarkably, despite the divergence of the vertex $\Lambda_\nu$ and the susceptibility $\chi_{\rm loc}$ at $T\rightarrow 0$, the vertex $\widetilde{\Lambda}_\nu$, which determines magnetic exchange, remains finite in this limit.
Performing the remaining integrations, we find the following contributions $J^{(2)}$ and $J^{(4)}=\sum_{i=1}^3 J^{(4)}_i$ to the magnetic exchange up to the order $t^2/U$ and $t^4/U^3$, respectively:
\begin{align}
J^{(2)} &=-2t_{ij}^{2}\int \frac{d\nu }{2\pi }\widetilde{\Lambda }_{\nu
}^{2}G_{\nu }^{4}=-\frac{4t_{ij}^{2}}{U},   \\
J_{1}^{(4)} &=-4t_{ij}t_{jk}t_{kl}t_{li} \int \frac{%
d\nu }{2\pi }\widetilde{\Lambda }_{\nu }^{2}G_{\nu }^{6}=\frac{4}{U^{3}}t_{ij}t_{jk}t_{kl}t_{li},  \nonumber\\
J_{2}^{(4)} &=-2
(t_{ik}t_{kj})(t_{il}t_{lj}) \int \frac{%
d\nu }{2\pi }\widetilde{\Lambda }_{\nu }^{2}G_{\nu }^{6}=\frac{2}{U^{3}}%
 (t_{ik}t_{kj})(t_{il}t_{lj}),  
\nonumber \\
J_{3}^{(4)} &=2t^2_{ik}t^2_{kj} T^2 \sum_{i\nu,i\nu'}\widetilde{\Lambda }_{\nu }G_{\nu }^{4}\widetilde{F}_{\nu
\nu ^{\prime }}G_{\nu ^{\prime }}^{4}\widetilde{\Lambda }_{\nu ^{\prime }}
=-\frac{6}{U^{3}}t^2_{ik}t^2_{kj} \notag
\end{align}
The contribution $J^{(2)}$ provides the well-known result for the negative (i.e. antiferromagnetic) nearest neighbor exchange interaction. In contrast to the previous derivation within the dual fermion approach \cite%
{Stepanov1}, we obtain it here from the paramagnetic, and not antiferromagnetic phase, which brings much closer similarity to the standard $t/U$ expansions. 

Summations over site indices in the contributions $J_2^{(4)}$ and $J_3^{(4)}$ reveal that for the square lattice both terms contribute to the next- and next-to-next nearest neighbor exchange interactions, for which we find the result $J_{2n}=J_{3n}=-4t^4/U^3$, coinciding with that obtained earlier within the standard $t/U$ expansion in Refs. \cite{Takahashi,McDonald}. Although the term $J_1^{(4)}$ provides a contribution $8t^4/U^3$ to the nearest-neighbor exchange interaction from cycling hoppings, which also coincides with that obtained previously in Refs. \cite{Takahashi,McDonald}, full determination of $t^4/U^3$ correction to the nearest neighbor exchange interaction within the presented scheme is rather cumbersome, since it requires also account of $t^2/U$ corrections to the local self-energy $\Sigma_\nu$ and the vertex $\widetilde{\Lambda}_\nu$, which correct $J^{(2)}$ at the order $t^4/U^3$. While the respective correction to the local self-energy was obtained previously \cite{Senechal}, calculation of corrections to the vertex $\widetilde{\Lambda}_\nu$  requires obtaining the $t^2/U$ correction to the interaction vertex $F_{{\rm loc},\nu\nu'}$, which were not, to our knowledge, considered earlier. Moreover, the non-local self-energy contribution, which is of the order $t^3/U^2$ (see, e.g., Ref. \cite{Senechal}), also provides a contribution to the nearest neighbor exchange interaction at the order $t^4/U^3$. Our preliminary study shows that the latter contribution, which is beyond the DMFT approach, is however to a large extent cancelled by the respective corrections to the RPA expression (\ref{Jqdef}). Detailed investigation of these issues is beyond of the scope of the present paper and postponed to future study.


\begin{thebibliography}{99}
\bibitem{goodenough} J. B. Goodenough, \textit{Magnetism and the Chemical
Bond} (Interscience, New York, 1963).

\bibitem{vonsovsky} S. V. Vonsovsky, \textit{Magnetism} (Wiley, New York,
1974).

\bibitem{mattis} D. C. Mattis, \textit{The Theory of Magnetism}
(Springer-Verlag, Berlin, 1981).

\bibitem{yosida} K. Yosida, \textit{Theory of Magnetism} (Springer-Verlag,
Berlin, 1996).

\bibitem{white} R. M. White, \textit{Quantum Theory of Magnetism: Magnetic
Properties of Materials}, 3rd ed. (Springer-Verlag, Berlin, 2007).


\bibitem{LKG1984} A. I. Liechtenstein, M. I. Katsnelson, and V. A. Gubanov, Exchange interactions and spin-wave stiffness in ferromagnetic metals, \href{https://dx.doi.org/10.1088/0305-4608/14/7/007}{J. Phys. F \textbf{14}, L125 (1984)}.

\bibitem{LKAG1987} A. I. Liechtenstein, M. I. Katsnelson, V. P. Antropov,
and V. A. Gubanov, 
Local spin density functional approach to the theory
of exchange interactions in ferromagnetic metals and alloys
\href{https://doi.org/10.1016/0304-8853(87)90721-9}{J. Magn. Magn. Mater. \textbf{67}, 65 (1987)}.

\bibitem{KL2000} M. I. Katsnelson and A. I. Lichtenstein, First-principles calculations of magnetic interactions in correlated systems \href{https://doi.org/10.1103/PhysRevB.61.8906}{Phys. Rev. B 
\textbf{61}, 8906 (2000)}.

\bibitem{NNNFe1} M. Pajda, J. Kudrnovsk\'y, I. Turek, V. Drchal, and P.
Bruno, Ab initio calculations of exchange interactions, spin-wave stiffness constants, and Curie temperatures of Fe, Co, and Ni,
\href{https://doi.org/10.1103/PhysRevB.64.174402}{Phys. Rev. B \textbf{64}, 174402 (2001)}.

\bibitem{Kats2004} M. I. Katsnelson and A. I. Lichtenstein, Magnetic susceptibility, exchange interactions and spin-wave spectra in the local spin density approximation, \href{https://dx.doi.org/10.1088/0953-8984/16/41/023}{J. Phys.:
Condens. Matter \textbf{16}, 7439 (2004)}.


\bibitem{gFe} A. V. Ruban, M. I. Katsnelson, W. Olovsson, S. I. Simak, and I. A. Abrikosov, 
Origin of magnetic frustrations in Fe-Ni 
Invar alloys, \href{https://doi.org/10.1103/PhysRevB.71.054402}{Phys. Rev. B \textbf{71}, 054402 (2005)}.

\bibitem{NNNFe2} Y. O. Kvashnin, R. Cardias, A. Szilva, I. Di Marco, M. I.
Katsnelson, A. I. Lichtenstein, L. Nordstr\"om, A. B. Klautau, and O.
Eriksson, 
Microscopic Origin of Heisenberg and Non-Heisenberg Exchange Interactions in Ferromagnetic bcc Fe, \href{https://doi.org/10.1103/PhysRevLett.116.217202}{Phys. Rev. Lett. \textbf{116}, 217202 (2016)}.

\bibitem{Solovyev2021} I. V. Solovyev, Exchange interactions and magnetic force theorem, \href{https://doi.org/10.1103/PhysRevB.103.104428}{Phys. Rev. B \textbf{103}, 104428 (2021)}.

\bibitem{chi} O. Grotheer, C. Ederer, and M. Fähnle, Fast ab initio methods for the calculation of adiabatic spin wave spectra in complex systems, 
\href{https://doi.org/10.1103/PhysRevB.63.100401}{Phys. Rev. B {\bf 63}, 100401(R) (2001)}. 

\bibitem{Antropov1} V. P. Antropov, The exchange coupling and spin waves in metallic magnets: removal of the long-wave approximation, \href{https://dx.doi.org/10.1016/S0304-8853(03)00206-3}{J. Magn. Magn. Mater. \textbf{262} L192 (2003)}

\bibitem{Bruno} P. Bruno, Exchange Interaction Parameters and Adiabatic Spin-Wave Spectra of Ferromagnets: A “Renormalized Magnetic Force Theorem”, \href{https://doi.org/10.1103/PhysRevLett.90.087205}{Phys. Rev. Lett. \textbf{90}, 087205 (2003)}.

\bibitem{Antropov} V. P. Antropov, No new ``Renormalized Magnetic Force Theorem", \url{https://arxiv.org/abs/cond-mat/0407739}; Magnetic short range order and the exchange coupling in magnets, \href{https://dx.doi.org/10.1016/j.jmmm.2005.10.220}{J. Magn. Magn. Mater. \textbf{300} e574 (2006)}.

\bibitem{DMFT_rev} A. Georges, G. Kotliar, W. Krauth and M. J. Rozenberg, Dynamical mean-field theory of strongly correlated fermion systems and the limit of infinite dimensions, \href{https://doi.org/10.1103/RevModPhys.68.13}{Rev. Mod. Phys. \textbf{68}, 13 (1996)}.

\bibitem{anisimov1997} V. I. Anisimov, A. I. Poteryaev, M. A. Korotin, A. O.
Anokhin, and G. Kotliar, First-principles calculations of the electronic structure and spectra of strongly correlated systems: dynamical mean-field theory, \href{https://dx.doi.org/10.1088/0953-8984/9/35/010}{J. Phys.: Condens. Mat. \textbf{9}, 7359 (1997)}.

\bibitem{LK1998} A. I. Lichtenstein and M. I. Katsnelson, Ab initio calculations of quasiparticle band structure in correlated systems: LDA++ approach, \href{https://doi.org/10.1103/PhysRevB.57.6884}{Phys. Rev. B 
\textbf{57}, 6884 (1998)}.

\bibitem{KL1999} M. I. Katsnelson and A. I. Lichtenstein, LDA++ approach to the electronic structure of magnets: correlation effects in iron, \href{https://dx.doi.org/10.1088/0953-8984/11/4/011}{J. Phys.: Condens.
Mat. \textbf{11}, 1037 (1999)}.

\bibitem{DFTplusDMFT} G. Kotliar, S. Y. Savrasov, K. Haule, V. S. Oudovenko,
O. Parcollet, and C. A. Marianetti, Electronic structure calculations with dynamical mean-field theory, \href{https://doi.org/10.1103/RevModPhys.78.865}{Rev. Mod. Phys. \textbf{78}, 865 (2006)}.

\bibitem{AbinitioDGA} A. Galler, P. Thunström, P. Gunacker, J. M. Tomczak, and K. Held, Ab initio dynamical vertex approximation, \href{https://doi.org/10.1103/PhysRevB.95.115107}{Phys. Rev. B {\bf 95}, 115107 (2017)}.

\bibitem{AbinitioDGA1} A. Galler, J. Kaufmann, P. Gunacker, P. Thunström, J. M. Tomczak, K. Held, Towards ab initio calculations with the dynamical vertex approximation, \href{https://doi.org/10.7566/JPSJ.87.041004}{J. Phys. Soc. Jpn. {\bf 87}, 041004 (2018)}.

\bibitem{OurJq} A. A. Katanin,
A. S. Belozerov, A. I. Lichtenstein, and M. I. Katsnelson, Exchange interactions in iron and nickel: DFT +DMFT study in paramagnetic phase, \href{https://doi.org/10.1103/PhysRevB.107.235118}{Phys. Rev. B {\bf 107}, 235118 (2023)}.

\bibitem{MyCo} A. A. Katanin, DFT+DMFT study of exchange interactions in cobalt and their implications for the competition of hcp and fcc phases \href{https://doi.org/10.1103/PhysRevB.108.235170}{Phys. Rev. B {\bf 108}, 235170 (2023)}

\bibitem{Loon} Erik G. C. P. van Loon, Hugo U. R. Strand, Dual Bethe-Salpeter equation for the multi-orbital lattice susceptibility within dynamical mean-field theory, \href{https://doi.org/10.1103/PhysRevB.109.155157}{Phys. Rev. B {\bf 109}, 155157 (2024)}.

\bibitem{MyCrO2} A. A. Katanin, Magnetic properties of half metal from the paramagnetic phase: DFT+DMFT study of exchange interactions in CrO$_2$, \href{https://doi.org/10.1103/PhysRevB.110.155115}{Phys. Rev. B {\bf 110}, 155115 (2024)}.

\bibitem{Hund1} P. Werner, E. Gull, M. Troyer, and A. J. Millis, Spin Freezing Transition and Non-Fermi-Liquid Self-Energy in a Three-Orbital Model, \href{https://doi.org/10.1103/PhysRevLett.101.166405}{Phys. Rev. Lett. {\bf 101}, 166405 (2008)}.

\bibitem{Hund2} Z. P. Yin, K. Haule, and G. Kotliar, Kinetic frustration and the nature of the magnetic and paramagnetic states in iron pnictides and iron chalcogenides, \href{https://dx.doi.org/10.1038/NMAT3120}{Nat. Mater. {\bf 10}, 932 (2011)}.

\bibitem{Hund3} L. de' Medici, J. Mravlje, and A. Georges, Janus-Faced Influence of Hund’s Rule Coupling in Strongly Correlated Materials, \href{https://doi.org/10.1103/PhysRevLett.107.256401}{Phys. Rev. Lett. {\bf 107}, 256401 (2011)}.

\bibitem{Hund4} K. M. Stadler, G. Kotliar, A. Weichselbaum, J. von Delft, Hundness versus Mottness in a three-band Hubbard–Hund model: On the origin of strong correlations in Hund metals, \href{https://doi.org/10.1016/j.aop.2018.10.017}{Annals of Physics {\bf 405}, 365 (2019)}.

\bibitem{OurFe} A. A. Katanin, A. I. Poteryaev, A. V. Efremov, A. O. Shorikov, S. L. Skornyakov, M. A. Korotin, and V. I. Anisimov, Orbital-selective formation of local moments in $\alpha$-iron: First-principles route to an effective model, \href{https://doi.org/10.1103/PhysRevB.81.045117}{Phys. Rev. B {\bf 81}, 045117 (2010)}.

\bibitem{OurGamma} P. A. Igoshev, A. V. Efremov, A. I. Poteryaev, A. A. Katanin, V. I. Anisimov, Magnetic fluctuations and effective magnetic moments in $\gamma$-iron due to electronic structure peculiarities, \href{https://doi.org/10.1103/PhysRevB.88.155120}{Phys. Rev. B \textbf{88}, 155120 (2013)}.

\bibitem{Sangiovanni} A. Hausoel, M. Karolak, E. Sasioglu, A. Lichtenstein, K. Held, A. Katanin, A. Toschi, and G. Sangiovanni, Local magnetic moments in iron and nickel at ambient and Earth's core conditions, \href{https://doi.org/10.1038/ncomms16062}{Nature Communications {\bf 8}, 16062 (2017)}.

\bibitem{AnisimovCo} A. S. Belozerov and V. I. Anisimov, Electron Correlation Effects in Paramagnetic Cobalt \href{https://doi.org/10.1134/S0021364023601379}{JETP Lett. {\bf 117}, 854 (2023)}. 

\bibitem{Stepanov1} E. A. Stepanov, S. Brener, F. Krien, M. Harland, A. I.
Lichtenstein, and M. I. Katsnelson, Effective Heisenberg Model and Exchange Interaction for Strongly Correlated Systems, \href{https://doi.org/10.1103/PhysRevLett.121.037204}{Phys. Rev. Lett. \textbf{121}, 037204
(2018)}.

\bibitem{Stepanov2} E. A. Stepanov, S. Brener, V. Harkov, M. I. Katsnelson,
and A. I. Lichtenstein, Spin dynamics of itinerant electrons: Local magnetic moment formation and Berry phase, \href{https://doi.org/10.1103/PhysRevB.105.155151}{Phys. Rev. B \textbf{105}, 155151 (2022)}.

\bibitem{StepanovRev} A. Szilva, Y. Kvashnin, E. A. Stepanov, L. Nordstr\"{o}%
m, O. Eriksson, A. I. Lichtenstein, M. I. Katsnelson, Quantitative theory of
magnetic interactions in solids, \href{https://doi.org/10.1103/RevModPhys.95.035004}{Rev. Mod. Phys. {\bf 95}, 035004 (2023)}

\bibitem{Izyumov} Yu. A. Izyumov, F. A. Kassan-Ogly, and Yu. N. Skryabin, Diagram technique for spin-operators and its applications to some problems of ferromagnetism,
\href{https://doi.org/10.1051/jphyscol:1971125}{Journ. de Phys. Colloq. \textbf{32}, C1-86 (1971)}; \textit{Field Methods in
the Theory of Ferromagnetism }[in Russian], Nauka, Moscow, 1974; Yu. A.
Izyumov and Yu. N. Skryabin, \textit{Statistical mechanics of magnetically
ordered systems}, Springer, 1988.

\bibitem{IgoshevKatanin} P. A. Igoshev, A. V. Efremov, A. A. Katanin, Magnetic exchange in $\alpha$-iron from ab initio calculations in the paramagnetic phase, \href{https://doi.org/10.1103/PhysRevB.91.195123}{Phys.
Rev. B \textbf{91}, 195123 (2015)}.

\bibitem{BelozerovKat} A. S. Belozerov, A. A. Katanin, V. I. Anisimov, Momentum-dependent susceptibilities and magnetic exchange in bcc iron from supercell DMFT calculations, \href{10.1103/PhysRevB.96.075108}{Phys. Rev. B {\bf 96}, 075108 (2017)}

\bibitem{Otsuki} J. Otsuki, K. Yoshimi, H. Shinaoka, and Y. Nomura, Strong-coupling formula for momentum-dependent susceptibilities in dynamical mean-field theory, \href{https://doi.org/10.1103/PhysRevB.99.165134}{Phys.
Rev. B \textbf{99}, 165134 (2019)}.

\bibitem{RefNote} We emphasize that in the presence of SU(2) symmetry in paramagnetic phase $\chi _{\mathbf{q}}^{rr^{\prime }}=\langle\langle S^{+,r}_{\mathbf{q}} | S^{-,r}_{-\mathbf{q}}\rangle\rangle_{\omega=0}$ and $\chi ^{r}_{\mathrm{loc}}=\langle\langle S^{+,r}_{i} | S^{-,r}_{i}\rangle\rangle_{\omega=0}$ can be also defined as related to the transverse susceptibilities. We consider longitudinal susceptibilities since they account correctly for the effect of Hund interaction in case of the density-density approximation of Coulomb interaction in paramagnetic phase. Passing to the ordered phase requires consideration of transverse susceptibilities, and, therefore, using SU(2) symmetric interaction.
 
\bibitem{chiDMFT} T. Pruschke, Q. Qin, T. Obermeier, and J. Keller, Magnetic properties of the t - J model in the dynamical mean field theory, \href{https://dx.doi.org/10.1088/0953-8984/8/18/009}{J. Phys.: Condens. Matter {\bf 8},
3161 (1996)}.

\bibitem{QE} P. Giannozzi, {\it et. al.}, {Quantum ESPRESSO: a modular and open-source software project for quantum simulations of materials},  \href{https://doi.org/10.1088/0953-8984/21/39/395502}{J. Phys.: Condens. Matter {\bf 21}, 395502 (2009)}; Advanced capabilities for materials modelling with Quantum ESPRESSO, \href{https://doi.org/10.1088/1361-648X/aa8f79}{ibid. {\bf 29}, 465901 (2017)}; \url{https://www.quantum-espresso.org}.

\bibitem{PPP} G. Prandini, A. Marrazzo, I. E. Castelli, N. Mounet, and N. Marzari, Precision and efficiency in solid-state pseudopotential calculations, \href{https://dx.doi.org/10.1038/s41524-018-0127-2npj}{Comput. Mater. {\bf 4}, 72 (2018)}; \url{http://materialscloud.org/sssp}.

\bibitem{Wannier90} G. Pizzi, et. al., {Wannier90 as a community code: new features and applications}, \href{https://doi.org/10.1088/1361-648X/ab51ff}{J. Phys. Cond. Matt. {\bf 32}, 165902 (2020)}; \url{http://www.wannier.org}.

\bibitem{AMF} M. T. Czy\ifmmode \.{z}\else \.{z}\fi{}yk and G. A. Sawatzky, Local-density functional and on-site correlations: The electronic structure of La$_2$CuO$_4$ and LaCuO$_3$, \href{https://doi.org/10.1103/PhysRevB.49.14211}{Phys. Rev. B \textbf{49}, 14211 (1994)}.

\bibitem{My_BS} A. A. Katanin, Improved treatment of fermion-boson vertices and Bethe-Salpeter equations in nonlocal extensions of dynamical mean field theory, \href{https://doi.org/10.1103/PhysRevB.101.035110}{Phys. Rev. B \textbf{101}, 035110 (2020)}.


\bibitem{CT-QMC} A. N. Rubtsov, V. V. Savkin, and A. I. Lichtenstein, Continuous-time quantum Monte Carlo method for fermions, \href{ https://doi.org/10.1103/PhysRevB.72.035122}{Phys.
Rev. B \textbf{72}, 035122 (2005)}; P. Werner, A. Comanac, L. de Medici, M.
Troyer, and A. J. Millis, Continuous-Time Solver for Quantum Impurity Models, \href{https://doi.org/10.1103/PhysRevLett.97.076405}{Phys. Rev. Lett. \textbf{97}, 076405 (2006)}.

\bibitem{iQIST} Li Huang, Y. Wang, Zi Yang Meng, L. Du, P. Werner, and Xi Dai, {iQIST: An open source continuous-time quantum Monte Carlo impurity solver toolkit}, \href{http://dx.doi.org/10.1016/j.cpc.2015.04.020}{Comp. Phys. Comm. \textbf{195}, 140 (2015)}; Li Huang, {iQIST v0.7: An open source continuous-time quantum Monte Carlo impurity solver toolkit},  \href{http://dx.doi.org/10.1016/j.cpc.2017.08.026}{Comp. Phys. Comm. \textbf{221}, 423 (2017)}.





\bibitem{triqs} O. Parcollet, M. Ferrero, T. Ayral, H. Hafermann, I. Krivenko, L. Messio, and P. Seth, TRIQS: A Toolbox for Research on Interacting Quantum Systems \href{https://doi.org/10.1016/j.cpc.2015.04.023}{Comp. Phys. Comm. {\bf 196}, 398 (2015)}

\bibitem{ct-hyb} P. Seth, I. Krivenko, M. Ferrero, 
and O. Parcollet, TRIQS/CTHYB: A continuous-time quantum Monte Carlo hybridisation expansion solver for quantum impurity problems, \href{https://doi.org/10.1016/j.cpc.2015.10.023}{Comp. Phys. Comm. {\bf 200}, 274 (2016)}.

\bibitem{DFTTools} M. Aichhorn, L. Pourovskii, 
P. Seth, V. Vildosola,
M. Zingl, O. E. Peil, 
X. Deng, J. Mravlje, 
G. J. Kraberger, C. Martins, M. Ferrero, and
O. Parcollet, TRIQS/DFTTools: A TRIQS application for ab initio calculations of correlated materials, \href{https://doi.org/10.1016/j.cpc.2016.03.014}{Comp. Phys. Comm.
{\bf 204}, 200 (2016)}.

\bibitem{AnisimovSU2} V. I. Anisimov, A. S. Belozerov, A. I. Poteryaev, I. Leonov, Rotationally-Invariant Exchange Interaction: The Case of Paramagnetic Iron, \href{https://doi.org/10.1103/PhysRevB.86.035152}{Phys. Rev. B {\bf 86}, 035152 (2012)}; A. S. Belozerov, I. Leonov, V. I. Anisimov, Magnetism of iron and nickel from rotationally invariant Hirsch-Fye quantum Monte Carlo calculations, \href{https://doi.org/10.1103/PhysRevB.87.125138}{Phys. Rev. B {\bf 87}, 125138 (2013)}.

\bibitem{Kondo1} K. Wilson, The renormalization group: Critical phenomena and the Kondo problem, \href{https://doi.org/10.1103/RevModPhys.47.773}{Rev. Mod. Phys. {\bf 47}, 773 (1975)}.

\bibitem{Kondo2} V. I. Mel’nikov, Thermodynamics of the Kondo problem, \href{http://jetpletters.ru/ps/1326/article_20037.shtml} {Soviet Phys. JETP Lett. {\bf 35}, 511 (1982)}.

\bibitem{Kondo3} A. M. Tsvelick and P. B. Wiegmann, Exact results in the theory of magnetic alloys, \href{https://dx.doi.org/ https://doi.org/10.1080/00018738300101581}{Adv. Phys. 32,
453 (1983)}.

\bibitem{Kondo4} A. A. Katanin, Extracting Kondo temperature of strongly-correlated systems from the inverse local magnetic susceptibility, \href{https://doi.org/10.1038/s41467-021-21641-2}{Nature Communications {\bf 12}, 1433 (2021)}.

\bibitem{Kondo5} X. Deng, K. M. Stadler, K. Haule, S.-S. B. Lee, A. Weichselbaum, J. v. Delft, and G. Kotliar, Reply to: Extracting Kondo temperature of strongly-correlated systems from the inverse local magnetic susceptibility, \href{https://doi.org/10.1038/s41467-021-21643-0}{Nature Communications {\bf 12}, 1445 (2021)}.

\bibitem{Hafermann} H. Hafermann, C. Jung, S. Brener, M. I. Katsnelson, A. N. Rubtsov and A. I. Lichtenstein, Superperturbation solver for quantum impurity models, \href{https://dx.doi.org/10.1209/0295-5075/85/27007}{Europhys. Lett., {\bf 85}, 27007 (2009)}.

\bibitem{ToschiV} G. Rohringer, A. Valli, and A. Toschi, Local electronic correlation at the two-particle level, \href{http://dx.doi.org/10.1103/PhysRevB.86.125114}{Phys. Rev. B {\bf 86}, 125114 (2012)}.

\bibitem{Takahashi} M. Takahashi, Half-filed Hubbard model at low temperature, \href{https://dx.doi.org/10.1088/0022-3719/10/8/031}{J. Phys. C {\bf 10}, 1289 (1977)}.

\bibitem{McDonald} A. H. MacDonald, S. M. Girvin, and D. Yoshioka, t/U expansion for the Hubbard model, \href{https://doi.org/10.1103/PhysRevB.37.9753}{Phys. Rev. B {\bf 37}, 9753 (1988)}; ibid. \href{https://doi.org/10.1103/PhysRevB.41.2565}{{\bf 41}, 2565 (1990)}; ibid. \href{https://doi.org/10.1103/PhysRevB.43.6209}{{\bf 43}, 6209 (1991)}. 

\bibitem{Senechal} S. Pairault, D. S\'en\'echal, and A.-M. S. Tremblay, Strong-Coupling Expansion for the Hubbard Model, \href{https://doi.org/10.1103/PhysRevLett.80.5389}{Phys. Rev. Lett. {\bf 80}, 5389 (1998)}.













\end{thebibliography}
\end{document}